\documentclass[twocolumn,trackchanges]{aastex631}

\usepackage{array}
\usepackage{hyperref}

\begin{document}

\def\msun{M$_\odot$}
\def\iso#1{$^{#1}$}

\title{Stellar Interpretation of Meteoritic Data and PLotting for Everyone (SIMPLE): \\
Isotope Mixing Lines for Six Sets of Core-Collapse Supernova Models}


\author{Marco Pignatari}

\affiliation{Konkoly Observatory, HUN-REN Research Centre for Astronomy and Earth Sciences, H-1121 Budapest, Konkoly Thege M. \'ut 15-17, Hungary}

\affiliation{CSFK, MTA Centre of Excellence, Budapest, Konkoly Thege Mikl\'os út 15-17, H-1121, Hungary}

\affiliation{University of Bayreuth, BGI, Universitätsstraße 30, 95447 Bayreuth, Germany}

\author{Mattias Ek}

\affiliation{Institute for Geochemistry and Petrology, ETH Zürich, Zurich, Switzerland}

\author{Georgy V. Makhatadze}

\affiliation{Université Paris Cité, Institut de Physique du Globe de Paris, CNRS, 1 rue Jussieu, Paris 75005, France}

\affiliation{Centre for Star and Planet Formation (StarPlan), Globe Institute, Faculty of Health and Medical Sciences, University of Copenhagen, \O{}ster Voldgade 5-7, 1350 Copenhagen K, Denmark}

\affiliation{Konkoly Observatory, HUN-REN Research Centre for Astronomy and Earth Sciences, H-1121 Budapest, Konkoly Thege M. \'ut 15-17, Hungary}

\affiliation{CSFK, MTA Centre of Excellence, Budapest, Konkoly Thege Mikl\'os út 15-17, H-1121, Hungary}

\author{G\'abor G. Bal\'azs}

\affiliation{Konkoly Observatory, HUN-REN Research Centre for Astronomy and Earth Sciences, H-1121 Budapest, Konkoly Thege M. \'ut 15-17, Hungary}

\affiliation{CSFK, MTA Centre of Excellence, Budapest, Konkoly Thege Mikl\'os út 15-17, H-1121, Hungary}

\affiliation{ELTE E\"{o}tv\"{o}s Lor\'and University, Institute of Physics and Astronomy, Budapest 1117, P\'azm\'any P\'eter s\'et\'any 1/A, Hungary}

\author{Lorenzo Roberti}

\affiliation{Istituto Nazionale di Fisica Nucleare—Laboratori Nazionali del Sud (INFN—LNS), Via Santa Sofia 62, Catania, Italy}

\affiliation{Konkoly Observatory, HUN-REN Research Centre for Astronomy and Earth Sciences, H-1121 Budapest, Konkoly Thege M. \'ut 15-17, Hungary}

\affiliation{CSFK, MTA Centre of Excellence, Budapest, Konkoly Thege Mikl\'os út 15-17, H-1121, Hungary}

\affiliation{Istituto Nazionale di Astrofisica—Osservatorio Astronomico di Roma (INAF—OAR), Via Frascati 33, I-00040, Monteporzio Catone, Italy}

\author{James M. Ball}

\affiliation{Institute for Geochemistry and Petrology, ETH Zürich, Zurich, Switzerland}

\author{Borb\'ala Cseh}

\affiliation{Konkoly Observatory, HUN-REN Research Centre for Astronomy and Earth Sciences, H-1121 Budapest, Konkoly Thege M. \'ut 15-17, Hungary}

\affiliation{CSFK, MTA Centre of Excellence, Budapest, Konkoly Thege Mikl\'os út 15-17, H-1121, Hungary}

\affiliation{MTA-ELTE Lend{\"u}let "Momentum" Milky Way Research Group, Hungary}

\author{Alessandro Chieffi}

\affiliation{Istituto Nazionale di Fisica Nucleare—Sezione di Perugia (INFN), via A. Pascoli s/n, I-06125 Perugia, Italy}

\affiliation{Istituto Nazionale di Astrofisica—Istituto di Astrofisica e Planetologia Spaziali (INAF—IAPS), Via Fosso del Cavaliere 100, I-00133, Roma, Italy}

\affiliation{School of Physics and Astronomy, Monash University, VIC 3800, Australia}


\author{Chris Fryer}

\affiliation{Center for Nonlinear Studies, Los Alamos National Laboratory, Los Alamos, NM 87545 USA}

\author{Falk Herwig}

\affiliation{Astronomy Research Centre, Department of Physics \& Astronomy, University of Victoria, Victoria, BC V8W 2Y2, Canada}

\author{Chiara Incollingo}

\affiliation{Institute for Geochemistry and Petrology, ETH Zürich, Zurich, Switzerland}

\author{Thomas Lawson}

\affiliation{E. A. Milne Centre for Astrophysics, University of Hull, Cottingham Road, Kingston upon Hull, HU6 7RX}

\author{Marco Limongi}

\affiliation{Istituto Nazionale di Astrofisica—Osservatorio Astronomico di Roma (INAF—OAR), Via Frascati 33, I-00040, Monteporzio Catone, Italy}

\affiliation{Kavli Institute for the Physics and Mathematics of the Universe, Todai Institutes for Advanced Study, University of Tokyo, Kashiwa, 277-8583 (Kavli IPMU, WPI), Japan}

\affiliation{Istituto Nazionale di Fisica Nucleare—Sezione di Perugia (INFN), via A. Pascoli s/n, I-06125 Perugia, Italy}

\author{Thomas Rauscher}

\affiliation{Department of Physics, University of Basel, Klingelbergstr. 82, CH-4056 Basel, Switzerland}
\affiliation{Centre for Astrophysics Research, University of Hertfordshire, Hatfield AL10 9AB, United Kingdom}

\author{Maria Sch\"onb\"achler}

\affiliation{Institute for Geochemistry and Petrology, ETH Zürich, Zurich, Switzerland}

\author{Andre Sieverding}

\affiliation{Lawrence Livermore National Laboratory, P.O. Box 808, L-414, Livermore, California 94551, USA}

\author{Reto Trappitsch}

\affiliation{Laboratory for Biological Geochemistry, School of Architecture, Civil \& Environmental Engineering, \'Ecole Polytechnique F\'ed\'erale de Lausanne, GR C2 505,
Station 2, 1015 Lausanne, Switzerland}

\author{Maria Lugaro}

\affiliation{Konkoly Observatory, HUN-REN Research Centre for Astronomy and Earth Sciences, H-1121 Budapest, Konkoly Thege M. \'ut 15-17, Hungary}

\affiliation{CSFK, MTA Centre of Excellence, Budapest, Konkoly Thege Mikl\'os út 15-17, H-1121, Hungary}

\affiliation{ELTE E\"{o}tv\"{o}s Lor\'and University, Institute of Physics and Astronomy, Budapest 1117, P\'azm\'any P\'eter s\'et\'any 1/A, Hungary}

\affiliation{School of Physics and Astronomy, Monash University, VIC 3800, Australia}

\correspondingauthor{Marco Pignatari}
\email{marco.pignatari@csfk.org} 
\correspondingauthor{Mattias Ek}
\email{mattias.ek@eaps.ethz.ch}
\correspondingauthor{Maria Lugaro}
\email{maria.lugaro@csfk.org}



\begin{abstract}

Bulk meteorites and their inclusions exhibit, for many chemical elements,  isotopic variability produced by nucleosynthetic events in stars and supernovae before the formation of the Sun.
While the exact astrophysical origins of these variations are still a matter of debate, their identification provides insights on the environment of the Sun's birth and the formation of the Solar System. Here we present a new Python tool called SIMPLE (Stellar Interpretation of Meteoritic Data and Plotting for Everyone) designed to compare the isotopic composition of the ejecta from core-collapse supernovae (CCSNe) with meteoritic data. In the present version, the SIMPLE toolkit includes a dataset of 18 CCSN models, from 6 different published sets, with initial masses of 15, 20, and 25 M$_{\sun}$ and solar metallicity. SIMPLE is designed to easily extract the isotopic abundances predicted by each CCSN model for any elements and post-process them into the format needed to compare to the meteoritic data, therefore, facilitating their interpretation. As an example of how to use SIMPLE, we analyze the composition of the Ni isotopes in the 18 models and confirm that bulk meteorite Ni anomalies are compatible with material from the innermost Si-rich region of CCSN ejecta. Designed as a collaborative platform, SIMPLE is open-source and welcomes community contributions to enhance its development and dissemination for stellar nucleosynthesis and meteoritic studies. Future enhancements include addition of more model predictions and inclusion of mixing between different layers of supernova ejecta.

\end{abstract}

\keywords{}


\section{Introduction} \label{sec:inthttps://www.overleaf.com/project/647da05ddfe094080a81bd76ro}


Materials derived from Early Solar System bodies and dust reservoirs exhibit a nucleosynthetic isotope variability across many chemical elements. These include elements of vastly different nucleosynthetic heritage, such as alpha- \citep{simon09}, iron peak \citep{rotaru92} and neutron-capture elements \citep{dauphas02a}. The variability is typically revealed by isotopic analysis of bulk meteorites and their components, such as calcium-aluminium-rich inclusions (CAIs) \citep{niemeyer81}. The origin of the Solar System heterogeneity is elusive and several models have been proposed, which can be divided into two major groups, and may both apply depending on the element and the variation considered. The heterogeneity could have been a product of separation of the anomalous carriers based on their thermal \citep{trinquier09, ek20} and/or aerodynamic \citep{steele12, hutchison22} properties. Alternatively, it could have been imprinted from a heterogeneous parental molecular cloud \citep{nanne19} or directly by a nearby supernova \citep{qin11, brennecka13, iizuka25}. Some models propose a combination of the two to explain different trends observed in inner and outer Solar System \citep{vankooten16}. Studying the nucleosynthetic variability of the Solar System can help constraining processes such as planetesimal \citep{lichtenberg21} and planet \citep{akram16Moon,kruijer17, schiller18, burkhardt21} formation, dust migration \citep{vankooten21} and evolution \citep{ek20} in the protoplanetary disc and infall dynamics from the parental molecular cloud \citep{vankooten16, nanne19}. While further information can be found in many recent reviews \citep{qin16, dauphas16, yokoyama16, bermingham20, kleine20, kruijer20, mezger20, REVbermingham20, bermingham24, liu24REV, REVbizzarro25, tissot25review, Schonbachler25REV}, caution is advised as the field is developing at a fast pace.

One of the steps required to understand the variability and exploit it to study the early Solar System is the identification of the nucleosynthetic inventory carried by various material in the early Solar System and, especially, which components could have been distributed heterogeneously and why. This requires a detailed comparison between the meteoritic isotope data and the predictions from models of stellar nucleosynthesis. 
The common approach to compare the meteorite data to the predictions of stellar nucleosynthesis is to plot a mixing line, between the bulk Solar System and the stellar component in question, which should pass through the data when the stellar component is diluted by the bulk Solar System (assumed to be the terrestrial) isotopic composition. While mixing line calculations are simple, in the case of extremely different isotope compositions such as the case with the products of nucleosynthesis, the aforementioned internal normalization introduces non-linear effects and special care must be be taken to avoid creating artifacts in the data. 
The comparison task is further complicated for a number of reasons: including the specifics of data reduction employed in isotope studies of bulk meteorites, the fact that the measured isotopic ratios reflect the complex history of the chemical and nuclear reactions that altered the sample and its precursor material, as well as the many stellar and nuclear physics uncertainties carried by the model predictions. 

In relation to the laboratory data, the typical magnitude of the isotope effects induced by chemical reactions is of the order of \(10^{-3}\), while the resolved stellar nucleosynthetic effects on a bulk meteorite can be as low as \(10^{-6}\). As such, separating the effects caused by nucleosynthesis from those caused by chemical and other nuclear processes can be a non-trivial task. Most chemical processes affect isotope compositions in a mass-dependent manner, which can be routinely corrected for using mass-dependent fractionation laws, resulting in a mass-independent isotope composition \citep{russell78}.  Chemical mass-independent isotope fractionation is also possible, however, typically these effects are limited to a few chemical elements \citep{dauphas16}. In-situ nuclear reactions, such as radioactive decay and nuclear reactions induced by exposure to cosmic rays, result in mass-independent isotope effects as well and correction for these is sometimes needed too. 

In relation to the stellar models, the full availability of data from model predictions is  limited, especially in the case of core-collapse supernovae (CCSN). These represent the final explosion of massive stars 
\citep[with initial mass roughly $>$ 9-10 \msun, e.g.,][]{limongi:24}, and are of particular interest in relation to the comparison to meteoritic data for a number of reasons. The main astrophysical reasons are that (i) they derive from massive stars, therefore their progenitors have relatively short lifetimes (from a few to a few tens of Myr). Therefore, they could have exploded in the same molecular cloud, or even the same stellar cluster, where the Sun was born. (ii) They are observed to produce dust, some of which has been found in the meteoritic stardust inventory. Probably up to 30\% of all the pre-solar stardust originally present in the presolar nebula originated from these explosions \citep{hoppe:22}. This dust is a likely candidate to have trapped the nucleosynthetic anomalies of the ejecta of their parent stars and carried them into the early Solar System.

Moreover, CCSN ejecta carry the composition produced by a large variety of nuclear-burning processes, from signatures of the hydrostatic nucleosynthesis that happened in the stellar supernova progenitor to the signature of the following explosive nucleosynthesis. While this diversity should be fully exploited to try to explain the observations, an especially crucial aspect of the comparison is represented by the analysis of the composition of individual parts of the ejecta, carrying their specific isotopic signatures. The analysis of the composition of each ejected shell is needed because CCSN ejecta are not fully mixed (as in the case of the winds from the fully convective envelopes of asymptotic giant branch stars), therefore dust can form locally. Evidence of this is the existence of meteoritic stardust grains, whose composition 
typically carries the signature of limited mixing between one or two different parts of the CCSN ejecta
\citep[e.g.,][]{zinner14,pignatari13grains}. 
The comparison between the composition of different CCSN shells and meteoritic data, however, has been only sparsely explored \citep{steele12, hopp22}. This is because for this task the  abundance profiles versus ejected mass are needed, which are usually not published or not as easily accessible as the total integrated stellar yields. It follows that only a few models have been used for such an analysis so far: \citet{meyer95,rauscher02,woosleyheger07}, which cannot capture the progress made in the past two decades on our understanding of CCSN nucleosynthesis. 

Models uncertainties stemming from nuclear and stellar physics are still large, especially as detailed nucleosynthesis is mostly available from one-dimensional models for both the progenitor and the explosion, although major efforts have been made and are under way to improve the situation, see recent review by \citet{fryer26}, as well as, e.g., \citet{burrows21,boccioli:24} and references therein. Nevertheless, considering different sets of models computed independently can help to evaluate systematic uncertainties at least in a qualitative way and establish which features of CCSN nucleosynthesis are robust, i.e., potentially mostly depending on the nuclear physics properties of the burning that produce them, and which features are instead model dependent. 

To advance on this task and support the common effort of the comparison between data and models, we developed a python tool called SIMPLE, which currently reads 18 CCSN models of masses 15, 20, 25 \msun\ from 6 different published sets, spanning from the popular models by \citet{rauscher02} to more recent models published from 2016 to 2022. The code includes a routine for calculating the correlations between isotopic ratios predicted by the models as a function of the ejected mass, to be directly compared to the meteoritic observations. The code can be downloaded as a python package and therefore can also be modified by each user as needed. Meetings and collaborations are envisaged for users so that we can share developments.

We note that the models available in SIMPLE are all Fe-core-collapse CCSNe, i.e., none of them represent explosions from core collapse driven by electron captures (EC) in stars of lower masses ($\sim$ 9 \msun). While these EC supernovae (ECSNe) are a popular option to explain \iso{48}Ca excesses in bulk meteorites \citep[e.g.][]{dauphas14Ca48,nittler:18}, their existence and occurrence at solar metallicity, their potential for dust production, as well as the large nuclear uncertainties affecting their ejecta are currently strongly debated \citep[e.g.,][]{jones:19a, jones:19b, denhartogh22}. This type of supernovae will therefore be addressed in future works, together with developments on CCSN nucleosynthesis from three-dimensional simulations and improved nuclear and neutrino input physics. The SIMPLE tool developed and presented here is able to incorporate future and updated models.  

The paper is structured as follows: in Section~\ref{sec:method} we describe the structure and workings of the SIMPLE code; in Section~\ref{sec:models} we describe the 18 CCSN models currently readable by the code. In Section~\ref{sec:nickel} we present some example cases of the usage of the code focusing on the Ni isotopes. In Section~\ref{sec:final} we summarize and discuss our work and describe the related future work. 



\section{Methods} \label{sec:method}

In this section we describe the main properties of the SIMPLE code, which allows comparison of CCSN abundances with bulk meteoritic data. The code can be downloaded from \href{https://github.com/mattias-ek/SIMPLE}{GitHub}\footnote{https://github.com/mattias-ek/SIMPLE} and documentation and tutorials for the code are available \href{https://mattias-ek.github.io/SIMPLE/}{here}\footnote{https://mattias-ek.github.io/SIMPLE/}. The code is also available on the Python Package Index (\texttt{pip install simple-chetec}). The stellar input data is available separately on Zenodo (\href{https://doi.org/10.5281/zenodo.17962172}{zenodo.17962172}\footnote{https://zenodo.org/records/17962172}), where they can be updated as needed. Below, we briefly describe the code functionalities. 



{\bf Reading the tabulated abundances.} The files that contain the tables for each model with the abundance of all the isotopes just after the explosion (see Section~\ref{sec:models} for details) as a function of location in mass of the ejecta are provided at the link above on Zenodo.
Isotopic abundances for each isotope $i$ are usually provided by CCSN modelers in mass fractions $X_i$\footnote{which corresponds to $n_i \times m_i/\rho = N_i \times m_i/M$, where $m_i= A \times amu$ is the mass of species $i$, $n_i = N_i/V$, i.e., the total number of nuclei $N_i$ per unit volume, $\rho$ the total density, and $M$ the total mass. By definition, $\Sigma_i\, X_i=1$.}. 
These can be directly used to calculate the total yield for each isotope, via integrating the product of the mass fraction and the mass of each resolved shell, over the whole ejected mass. When divided by their mass, each isotopic mass fraction $X_i$ is converted into $Y_i=X_i/A_i$, the molar fraction of the isotope, or ``normalized'' abundance by number. The $Y_i$ and $X_i$ quantities are not equivalent to use for normalization: they should match the same quantity at which the solar/terrestrial/standard abundances are given. Otherwise, for any ratio of any isotope $i$ over $j$ a factor $A_i/A_j$ is wrongly introduced to the calculations. 
Moreover, calculation of radioactive decay and addition of possible chemistry fractionation effects (see below) must be done in molar fractions. Therefore, the code is designed to be flexible in converting the data from mass fractions to molar fractions and vice versa, using the approximated masses, i.e., using mass 56 for \iso{56}Ni instead of 55.942, etc. The code works self-consistently with either representation, as chosen by the user. To avoid confusion, an attribute is associated with the data ``mass'' for mass fraction and ``mole'' for the abundance. As detailed in Section~\ref{sec:models}, not all the elements and isotopes are available in all the models and the code will issue a warning in case of missing isotopes. Also implemented is the possibility to include, in each figure where the x-axis represents the mass coordinate, separation lines and labels indicating the zonal structure according to the most abundant burning products, following \citet{meyer95}. 


{\bf Solar/terrestrial/standard abundances.} To normalize stellar models predictions we cannot use the laboratory standards used to normalize the experimental data. Instead, stellar isotopic variations need to be calculated relative to their own ``standard'' abundances, i.e., the initial abundances used in each model. If different numbers are used, fictional anomalies may appear that are not due to nuclear burning but to the difference between the initial values used in the stellar model calculations and the different values used. As the initial abundances are different for each sets of models (Section~\ref{sec:models}), and their tabulated values include a different number of digits, 
the initial abundances used during the computation of each stellar model is automatically used as the standard abundance for all normalization procedures.


{\bf Radioactive nuclei.} 
While creating the abundance tables of selected isotopes to analyze, the user can decide, case by case, how to handle the abundances of radioactive nuclei into those of their stable daughter nuclei. The current version of the code does not include yet a radioactive decay routine, therefore, the abundances of radioactive nuclei can only be omitted completely or added in their entirety to the abundance of their stable daughter nuclei. Once an element or isotope system is selected to be studied by the user, a preliminary step for the analysis is to decide which radioactive isotopes should be added to each stable isotopes. The stellar abundances are given relatively soon after the explosion (at most after $\sim$7 hours, see Section~\ref{sec:models}), therefore, most radioactive nuclei need to be added if the user wishes to assume that they were incorporated into dust, given that dust formation timescales in CCSNe are of the order of months to years \citep{brooker22}. If a user requests a radioactive isotope that is not included in a model, the abundance of this isotope is automatically set to zero. An example is shown in Table~\ref{tab:decay}, which list the radioactive nuclei to be considered for the Ni isotopes. In this case, all the radioactive isotopes can be safely added to their respective stable daughter nucleus, given their short half lives, except for the case of \iso{60}Fe, which needs a more careful analysis because it may or may not be incorporated in the same dust as Ni. In this case it is interesting to compare the \iso{60}Ni abundance, with and without the inclusion of \iso{60}Fe, as done in Figure~\ref{fig:Ni60decayed_undecayed20}. Except for this figure, all the results presented in Section~\ref{sec:nickel} are calculated using abundances where all the radioactive isotopes are fully decayed. 


{\bf Dilution and internal normalization.} For comparison to the meteoritic data, the stellar ejecta need to be diluted, internally normalized, and expressed as $\varepsilon$ values, i.e., per 10,000 deviations relative to the initial stellar abundances, as described above \citep[see full derivation, e.g., in][]{lugaro23}. For example, in the case of $\varepsilon$\iso{64}Ni$^*_{(81)}$ in Figure~\ref{fig:Nislope20}, and \ref{fig:zoneplot}, the * is used just to indicate that all the radioactive isotopes are fully decayed (see previous section), and:


\begin{equation}
    \varepsilon ^{64}{\rm Ni}_{(81)} = {\left(
    \frac{(^{64}{\rm Ni}/^{61}{\rm Ni})^{norm}_{}}{(^{64}{\rm Ni}/^{61}{\rm Ni})^{init}_{}} - 1
    \right)} \times 10^4,
    \label{eq:epsilon}
\end{equation}

\noindent where $(^{64}{\rm Ni}/^{61}{\rm Ni})^{init}$ is the initial stellar ratio used in each model, and $(^{64}{\rm Ni}/^{61}{\rm Ni})^{norm}$ is the ratio internally normalized using $^{58}{\rm Ni}/^{61}{\rm Ni}$, i.e.:

\begin{equation}
    (^{64}{\rm Ni}/^{61}{\rm Ni})^{norm} = 
    (^{64}{\rm Ni}/^{61}{\rm Ni})^{dil}{\left(
    \frac{(^{58}{\rm Ni}/^{61}{\rm Ni})^{init}}{(^{58}{\rm Ni}/^{61}{\rm Ni})^{dil}}
    \right)}^{-Q}.
    \label{eq:intnorm}
       \end{equation}

\noindent Here, 

\begin{equation}
Q = \frac{\ln{(64)} - \ln{(61)}}{\ln{(58)} - ln{(61)}} \sim \frac{64 - 61}{58 - 61},
\label{eq:Q}
 \end{equation}

\noindent and $(^{64}{\rm Ni}/^{61}{\rm Ni})^{dil}$ and $(^{58}{\rm Ni}/^{61}{\rm Ni})^{dil}$ are the isotopic ratio derived after dilution of the stellar abundances with the solar abundances\footnote{here these also correspond to the initial stellar abundances, but they are conceptually different as they would need to be scaled if models of different metallicity are considered} and  using the free parameter $x$ as the dilution factor, i.e.,

\begin{equation}
    ^{64}{\rm Ni} = ^{64}{\rm Ni}_\odot +  \left(^{64}{\rm Ni}^{star} \times x\right).
    \label{eq:dil}
\end{equation}

This dilution (or linearisation) is necessary as the mixing lines between the internally normalized Solar System and the extremely anomalous compositions of the stellar nucleosynthetic predictions are not straight, as one might expect, due to the effects introduced by exponential law used for the normalization. These effects are only negligible in the case of the relatively similar mixing end-member isotope compositions.
To perform this transformation, the default method used by the code dilutes the stellar abundances such that the largest off-set from the model initial abundances among all the isotopic ratios calculated from all the selected isotopes, corresponds to an $\varepsilon$ value of $\pm 1$. This value can be adjusted by the user and the corresponding dilution factor can be extracted. A minimum value of the dilution factor can also be set, which helps to avoid spurious results that can occur when the abundances (for example, in the H-rich envelope where not much nuclear burning happened) are too close to the initial values. For comparison the linear approximation method from \citet{dauphas04} is also implemented in the code, including the choice of approximating the $Q$ value using masses instead of ln(masses) (see Eq.~\ref{eq:Q}). 

{\bf Chemical fractionation.} 
When using SIMPLE to plot isotopic ratios of two different elements against each other, it is possible to apply chemical enrichment factors to simulate chemical fractionation, i.e., elemental separation according to chemical properties, which changes the elemental relative abundances and therefore the relative dilution. This can be done in two ways: either the abundances can be modified using enrichment factors as multiplication factors to the stellar abundances; or, the relative abundances of chemical elements can be set to given, fixed values (e.g., corresponding to the relative proportions of Mg and Al in spinel).

{\bf Validation test.} To check that the code works properly, we compared the results for the Ni isotopes to those published by \citet{steele12} in their Figure 9 (top panels), using the 25 \msun\ model from \citet{rauscher02}, which is also included in our dataset. We found that our results are in agreement with those published only when using a different model from the \citet{rauscher02}'s set than the ``a28''-labelled model included in our dataset,  
which was calculated with the standard choice of the \iso{22}Ne($\alpha$,n)\iso{25}Mg reaction rate defined in that paper. Other small differences also present are attributable to the use of different abundance quantities and to spurious results in the stellar envelope due to the choice of the solar/terrestrial abundances (see discussion above). 

\section{Models} \label{sec:models}

The SIMPLE package includes 6 sets of CCSN models, each including 3 stars of initial mass 15, 20, and 25 \msun\ and solar metallicity, for a total of 18 models. Out of these sets of models, only one \citep{rauscher02} has so far been extensively used for comparison to data from meteoritic bulk rocks and inclusions. The other model mostly used so far has been the 25 \msun\ model published by \citet{meyer95}, which was calculated using the same stellar evolution code and explosion mechanism as the \citet{rauscher02}'s models (see details in the next subsection). \tablename~\ref{tab:models} lists the different sets and their acronyms in alphabetical order. The main features of each set are described in the following subsections - listed in chronological order as references are often needed from one set to the following. 
All the stellar progenitors are non-rotating stars and the exact value of the solar metallicity and the initial isotopic distribution depends on the set, and are reported in detail below. The abundances are provided soon after the explosion, the exact time depending on the set: in the LC18, Ra02, and Si18 sets the time is set to 25000 s ($\sim 7$ hours), while in the other models it is set as the time when the peak temperature decreases below approximately 1.5 GK, after which there is not more nucleosynthesis. Note that also these times are very short, and much shorter than the timescale of dust formation probably of the order of years \citep{brooker22}. As mentioned above, care is therefore needed when deciding which radioactive abundances to add to the stable isotopes for each different element. 

\subsection{The Ra02 set}

The Ra02 set was extracted from \cite{rauscher02}, where we selected the models corresponding to the publication, i.e., those labelled as S15, S20, S25 in the publication and s15a28c, s20a28n, and s25a28d in the online database (nucleosynthesis.org, noting that the files from the first table called expl\_comp are required for the self-consistent treatment of the radioactive decay between the code and the files). The stellar progenitors were calculated with the Kepler code \citep{weaver:78} adopting an initial metallicity of Z=0.02 \citep{anders89}. The explosion was simulated by the piston-driven mechanism, where a piston is placed outside the iron core and its trajectory is calibrated by requiring a fixed kinetic energy in the ejecta measured at infinity. In the 15 \msun\ model, this energy is 1.2 foe (1 foe = $10^{51}$ erg), which corresponds to that estimated for SN 1987A. In the 20 and 25 \msun\ models, instead, the energy is fixed to 2.2 and 1.7 foe, respectively, to produce 0.1 \msun\ of \iso{56}Ni. The remnant mass (i.e., below the so-called ``mass-cut'') for each model is determined by the solution of the piston model. The nucleosynthesis is calculated during the evolution at the end of each timestep with a large adaptive network including up to 2200 nuclear species. Note that the 20 \msun\ model of this set experiences a C-O shell merger \citep{ritter18,roberti:23,roberti25} shortly before the explosion, which affects its structure and nucleosynthesis.

\subsection{The Pi16 set}

The Pi16 set includes the models published in \citet{pignatari:16} and computed using the GENEC code \citep{eggenberger:08}, up to the end of central Si burning. The initial metallicity is Z=0.02 \citep{grevesse:93}, with the isotopic distribution given by \cite{lodders03}.
The explosion was simulated using a semi-analytical approach, with the mass-cut prescription from \citet{fryer12} and the \citet{sedov:46} blast wave (SBW) solution for the determination of the peak velocity of the shock. The initial shock velocity for each model is $2\times10^9\ \rm cm/s$.
The maximum velocity of the shock front is capped at $5\times10^9\ \rm cm/s$ to mimic deceleration due to viscous forces. The adiabatic cooling of the shocked material is simulated using a variant of the adiabatic exponential decay \citep{hoyle:64,fowler:64}. The nucleosynthesis during both the evolution and the explosion is calculated by post-processing the previously computed stellar structure using the Multi-zone Post-Processing Network -- Parallel code \citep[MPPNP,][from the NuGrid collaboration\footnote{\url{www.nugridstars.org}}]{Pignatari:2012dw}, using a dynamical network that includes more than 5000 nuclear species and more than 70,000 reactions. Note that the 25 \msun\ model of this set has a rather poor resolution in mass in the O/C layer (see Section \ref{sec:nickel} for the definition) and care is needed when considering this model. 

\subsection{The Si18 set}

The Si18 set from \cite{sieverding:18} is based on stellar models calculated using the Kepler code like the Ra02 set. 
The explosion is simulated by the piston-driven mechanism, imposing 1.2 foe of kinetic energy in the ejecta measured at infinity in all the three models. The nucleosynthesis is calculated as in Ra02. At variance with all the other sets, the Si18 models include also a number of updated $\nu$--induced reactions. Since the $\nu$--induced nucleosynthesis depends on the neutrino energy, \cite{sieverding:18} presented a parametric study of the explosive yields exploring several neutrino energies. The models included in our database are those with the highest neutrino energy, the same used in \cite{denhartogh:22} and \cite{roberti:23}.

\subsection{The Ri18 set}

The Ri18 set includes the models published in \cite{ritter18set}, for which the progenitors were computed using the publicly available MESA code \cite[e.g.,][]{paxton:11}. As in Pi16, the initial metallicity is Z=0.02 \citep{grevesse:93}, with the isotopic distribution given by \cite{lodders03}. The explosions are performed as in Pi16, with the same semi-analytical approach and using the same initial shock velocity of $2\times10^9\ \rm cm/s$. The nucleosynthesis during both the evolution and the explosion was calculated by post-processing using the MPPNP code. Note that, like the 20 \msun\ model of Ra02, the 15 \msun\ model of this set experiences a C-O shell merger shortly before the explosion. The signatures of these events will be described in Section~\ref{sec:nickel}.

\subsection{The LC18 set}

The LC18 set includes the recommended (R) models from \citet{limongi18}, calculated with the Frascati Raphson Newton Evolutionary Code (FRANEC). The initial metallicity is Z=0.01345 \citep{asplund09}. While the full database\footnote{Online Repository for the Franec Evolutionary Output (ORFEO): \url{https://orfeo.oa-roma.inaf.it}} also includes rotating models with initial rotational velocity of 150 and 300 km s$^{-1}$, here we selected only the non-rotating models. The explosion is simulated as a kinetic bomb, i.e., by injecting a certain amount of kinetic energy into the Fe core and then following the propagation of the generated shock by solving the hydrodynamic equations. While the choice of the injected energy is made to obtain the complete ejection of all the stellar layers above the Fe core, these models are also set to experience a so-called mixing and fall-back \citep[M$\&$F,][]{umeda02}. The inner border of the mixed region is chosen by imposing that Ni/Fe = 1.6 (Ni/Fe)$_{\odot}$, and the outer border by imposing an \iso{16}O mass fraction below 0.1, i.e., the mixed region cannot extend beyond the base of the oxygen shell. The amount of fall-back material is then determined by requiring an ejection of 0.07 \msun\ of \iso{56}Ni. 
The nucleosynthesis is calculated during the evolution and the isotopic composition is fully coupled to the solution of the physics and mixing equations. The nuclear network includes 335 nuclear species and more than 3000 nuclear reactions, both for the progenitor evolution and the explosion. The isotopes included range from H to \iso{209}Bi. Up to \iso{98}Mo, all the possible weak and strong reactions connecting the various nuclear species are included. Above \iso{98}Mo, only the neutron-capture nucleosynthesis is considered, i.e., only (n,$\gamma$) reactions and $\beta^{-}$ decays, and only the stable and unstable isotopes around the magic numbers of neutrons N=82 and N=126 are explicitly included, while all the other intermediate isotopes are assumed to be at local equilibrium.

\subsection{The La22 set}

The La22 set includes models published in \citet{jones:19}, \citet{andrews:20} and \citet{lawson22}, which were computed using a more recent version of the Kepler code \citep{heger10} compared to the progenitors models of the Ra02 and Si18 sets. Like in Pi16 and Ri18, Z=0.02 \citep{grevesse:93}, and the isotopic distribution is from \citet{lodders03}. The explosions are modeled using a 1D hydrodynamic code that simulates the effects of the 3D convection-enhanced supernova engine of \cite{fryer18}, exploring various explosion parameters such as power, duration, and the extent of the energy injection region within the stellar progenitors. The complete dataset includes over 60 combinations of these parameters; while 
the models currently included in the SIMPLE toolkit are the same as those used by \citet{denhartogh:22} and \citet{roberti:23}. These correspond, for each initial mass, to the models with the explosion energy closest to the value of 1.2 foe used by Si18. Like for Pi16 and Ri18, the progenitor nucleosynthesis was calculated by post-processing using the MPPNP code. The explosive nucleosynthesis was instead calculated using a different post-processing code, the Tracer particle Post-Processing Network -- Parallel code \citep[TPPNP,][]{jones:19}, with the same nuclear network used for MPPNP.






\begin{table*}
\begin{center}
\footnotesize
\caption{Main methods and features of the six sets of CCSN models available in the current SIMPLE toolkit, see text for details. As in the SIMPLE code, the labels stand for: La22 = \citet{lawson22}; LC18 = \citet{limongi18}; Pi16 = \citet{pignatari16sets}; Ra02 = \citet{rauscher02}; Ri18 = \citet{ritter18set}; Si18 = \citet{sieverding:18}. The sets are listed here in alphabetical order and in the text in historical order. The time after explosion is the time at which the abundances are given in the tables read by SIMPLE.
\label{tab:models}}
\begin{tabular}{ccccccc} 
\hline
Set & La22 & LC18 & Pi16 & Ra02 & Ri18 & Si18 \\
\hline
\multicolumn{7}{c}{Progenitor models}\\
\hline
Solar distribution$^{a}$   & GN93/L03 & A09 & GN93/L03 & AG89 & GN93/L03 & L03 \\
Evolution code & Kepler & FRANEC & GENEC & Kepler & MESA & Kepler \\
Nucleosynthesis code & MPPNP & FRANEC$^{b}$ & MPPNP & Kepler$^{b}$ & MPPNP & Kepler$^{b}$ \\
Nuclear species & 5000+ & 300+ & 5000+ & 2200+ & 5000+ & 2200+ \\
\hline
\multicolumn{7}{c}{Supernova models}\\
\hline
Explosion engine & $\nu$-driven  & Kinetic bomb & Sedov blast-wave & Piston & Sedov blast-wave & Piston \\
Mass Cut & Hydro-based & $^{56}$Ni=0.07\msun & Fryer+12 & $E_{\rm exp}$=1.2 foe & Fryer+12 & $E_{\rm exp}$=1.2 foe \\
        & & with M$\&$F$^c$ & & (for 15 \msun) & & \\
         &             &          &          & $^{56}$Ni = 0.1 \msun & & \\
                  &             &          &          & (for 20, 25 \msun) & & \\

Remnant mass$^d$ (\msun) & 1.74, 2.27, 1.83 & 1.67, 1.66, 1.96 & 1.63, 2.81, 5.77 & 1.68, 1.55, 2.04 & 1.61, 2.74, 5.72 & 1.54, 1.58, 2.10 \\
Nucleosynthesis code & TPPNP & FRANEC$^{b}$ & MPPNP & Kepler$^{b}$ & MPPNP & Kepler$^{b}$ \\
Time after explosion (s) & $<$200 & 25000 & $<$200 & 25000 & $<$200 & 25000  \\
 \hline
       \end{tabular}
\end{center}      
$^a$Where GN93/L03 = \cite{grevesse:93} with the isotopic distribution of \cite{lodders03}; A09 = \citet{asplund09}; AG89 = \citet{anders89}.  
All metallicities $Z$ = 0.02, except for LC18 which uses $Z$ = 0.01345. \\
$^b$The nucleosynthesis is calculated together with the structure, either simultaneously with a fixed network (FRANEC) or after each time step with an adaptive network (Kepler).\\
$^c$Mixing and fall-back method, see text for details. \\
$^d$For the 15, 20, and 25 \msun\ models, respectively.

\end{table*}

\section{Nickel as a case study} \label{sec:nickel}

In this section, we show how to use SIMPLE to data-mine stellar models and compare the models predictions with meteoritic data. We focus on the production of the Ni isotopes in the 20 M$_{\odot}$ models. Analogous figures for the 15 M$_{\odot}$ and 25 M$_{\odot}$ models are also available in a Zenodo repository\footnote{https://zenodo.org/records/16743238}, together with the jupyter notebooks and the support files needed to generate them using the SIMPLE package.
Here, we use the approach common in the literature of comparing CCSN models generated from stellar progenitors of the same initial mass \citep[e.g.,][]{denhartogh:22}, 
however, we highlight that the CO core mass at the end of central He burning, which depends on the evolution of the star up to that stage, is a more reliable predictor of subsequent evolution and nucleosynthesis than the initial total stellar mass \citep[e.g.,][]{woosley95, thielemann:96, chieffi98,limongi18}. For example, a 20 M$_{\odot}$ model in a given set could have a similar CO core mass as a 17 M$_{\odot}$ model in another set, making its behavior more comparable to the latter, than to a 20 M$_{\odot}$ model from the same set. Using the CO core mass as a reference would therefore help to mitigate some of the discrepancies observed here between different models. To achieve a more consistent comparison, finer grids of stellar models should be developed in future studies.

\begin{figure*}
     \includegraphics[width=0.5\linewidth]{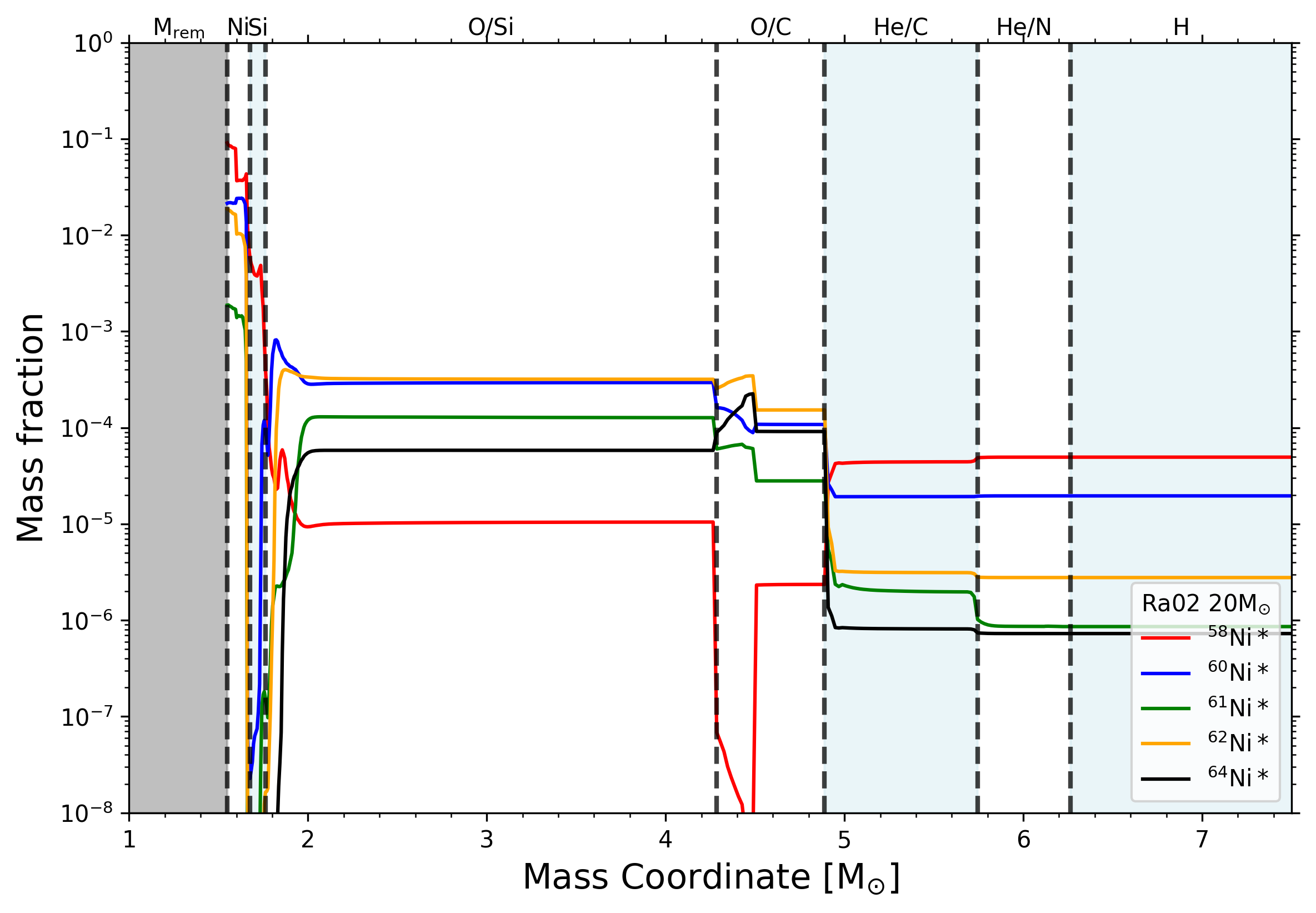}
     \includegraphics[width=0.5\linewidth]{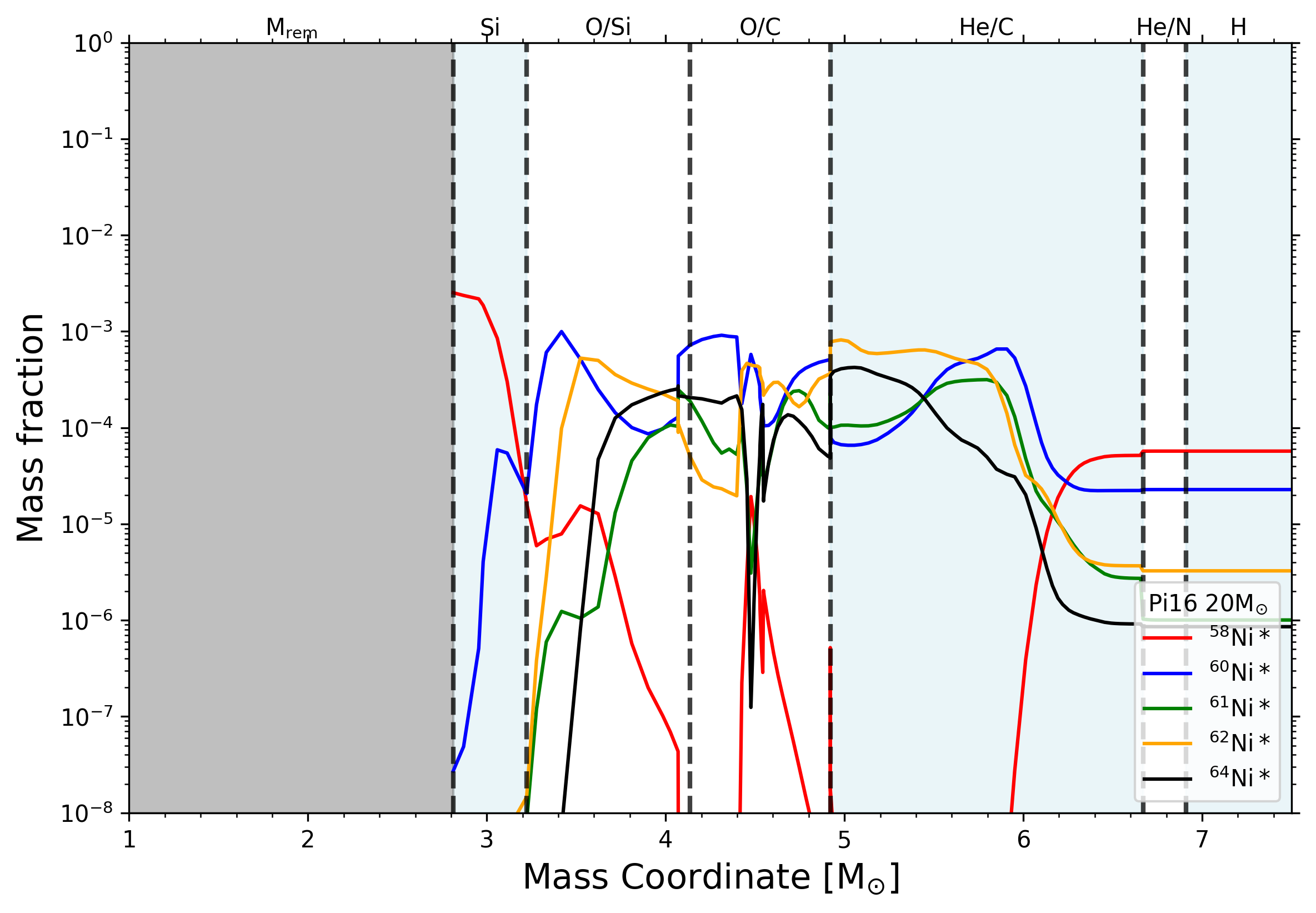}
     \includegraphics[width=0.5\linewidth]{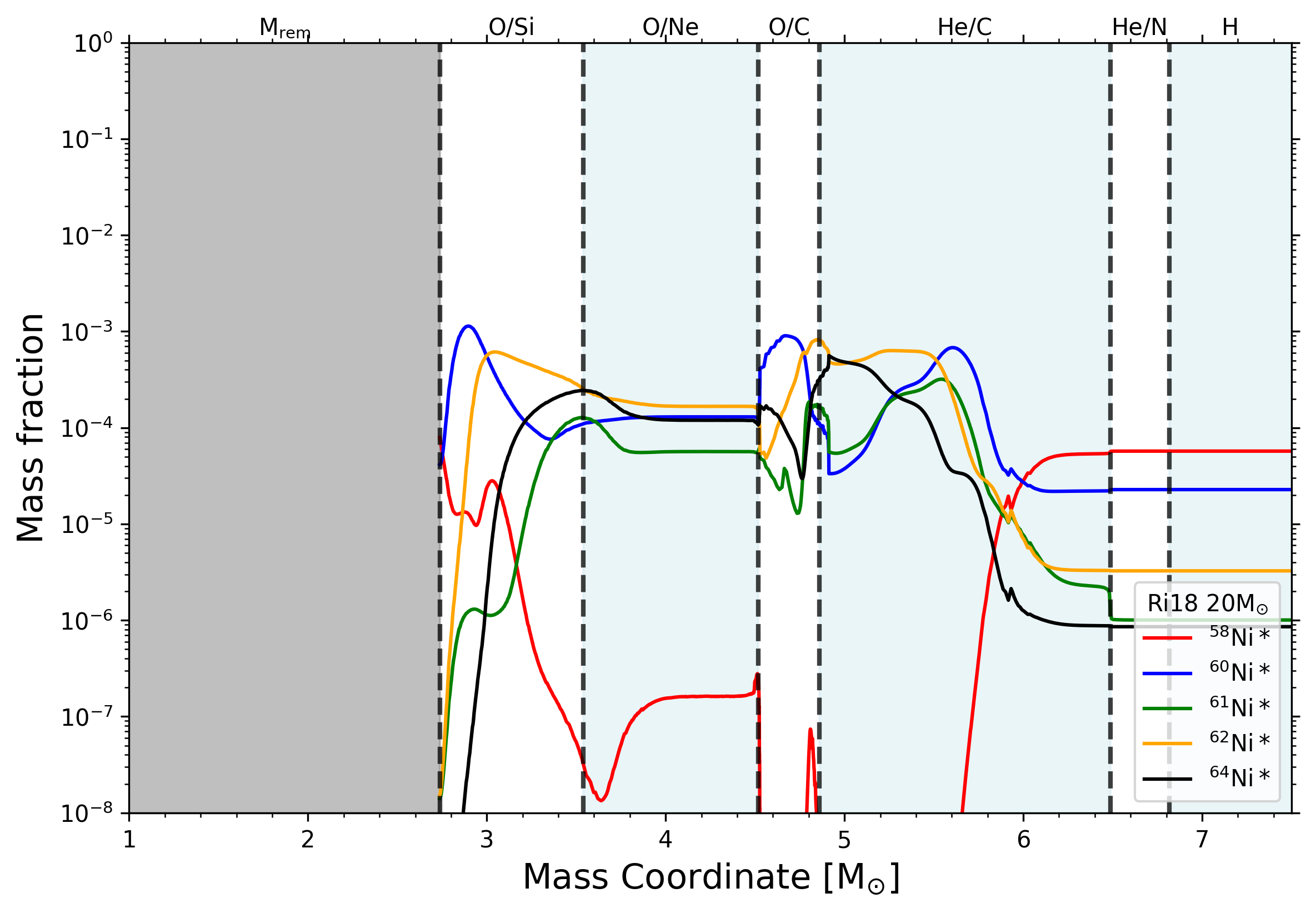}
     \includegraphics[width=0.5\linewidth]{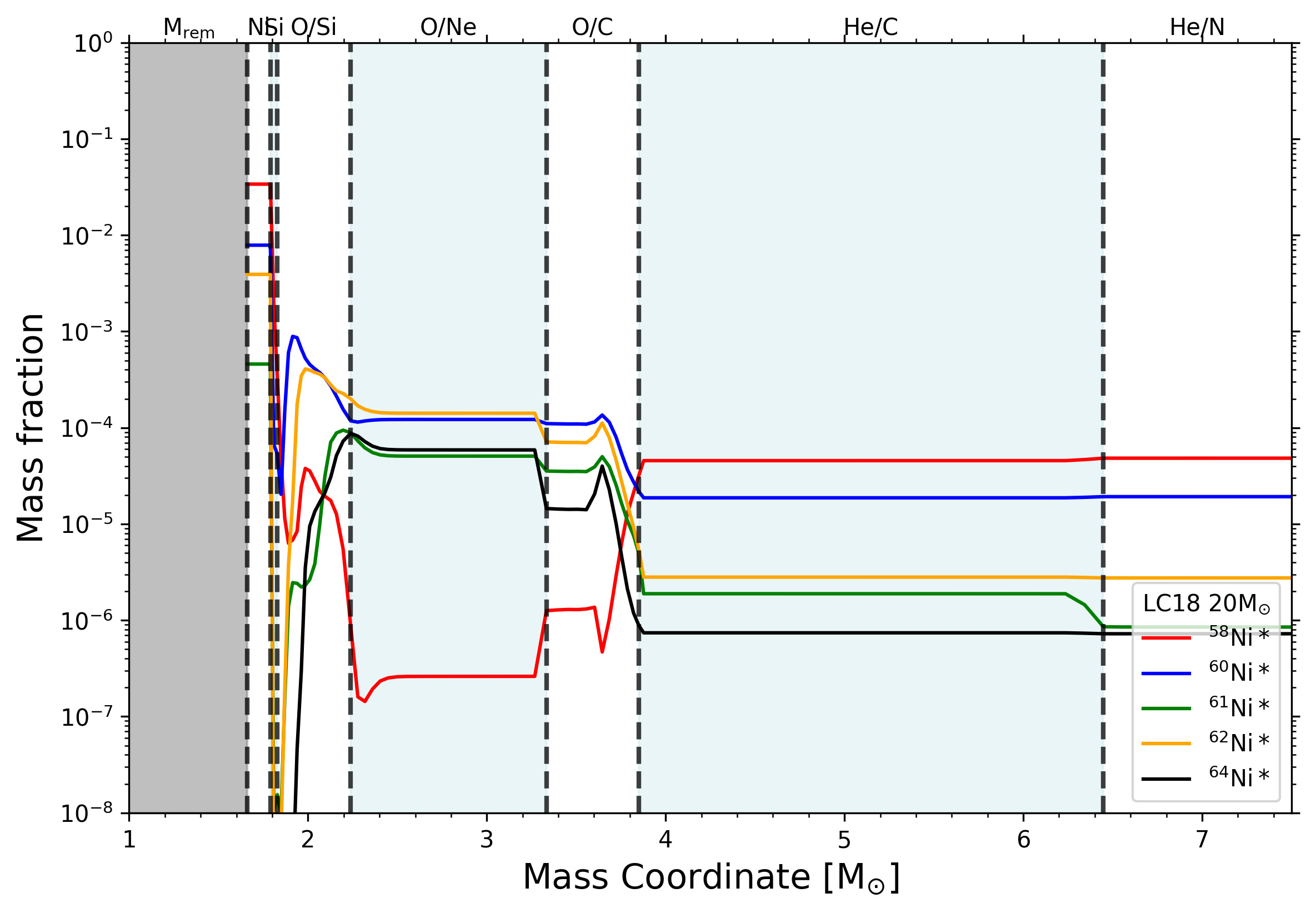}
     \includegraphics[width=0.5\linewidth]{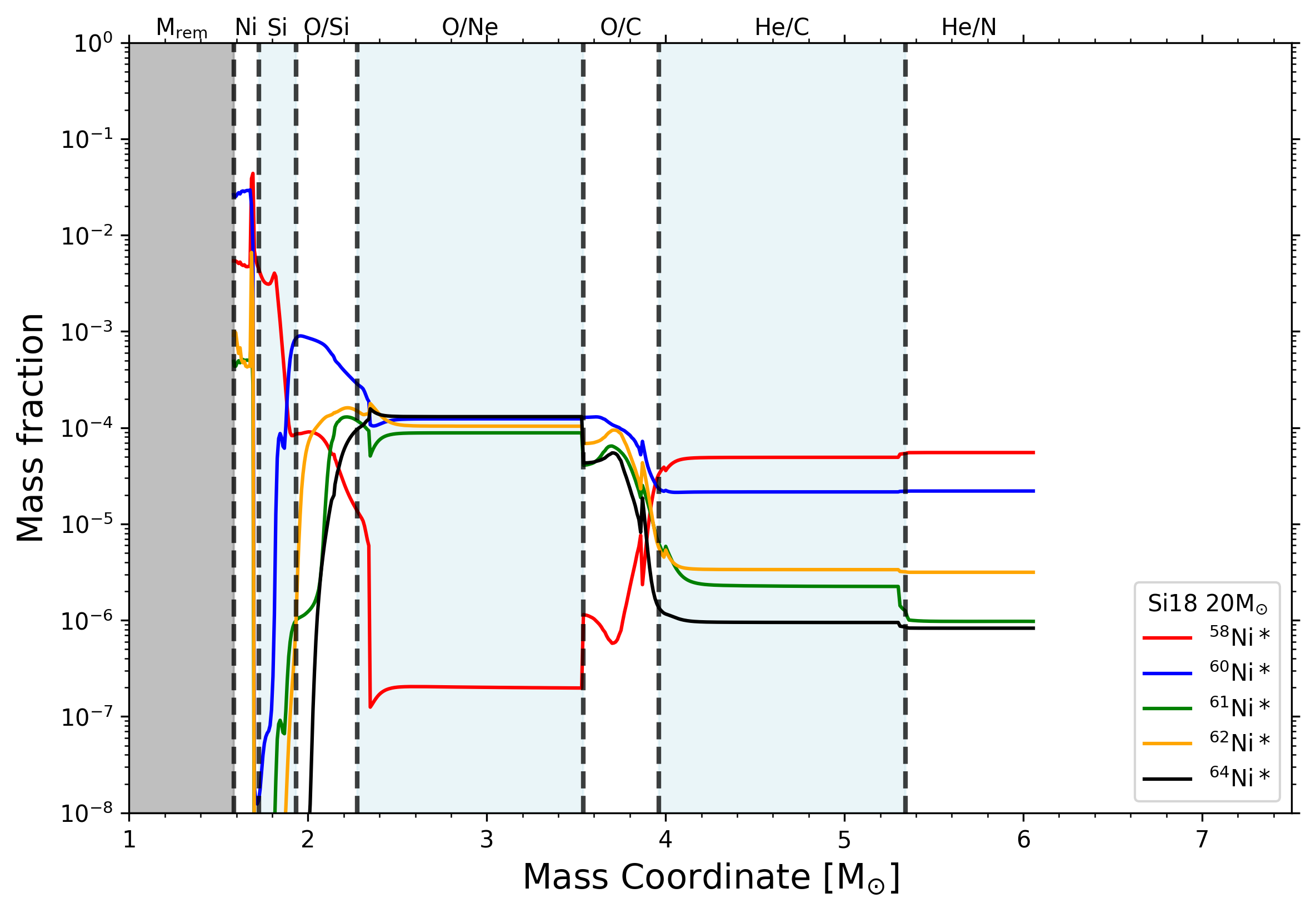}
     \includegraphics[width=0.5\linewidth]{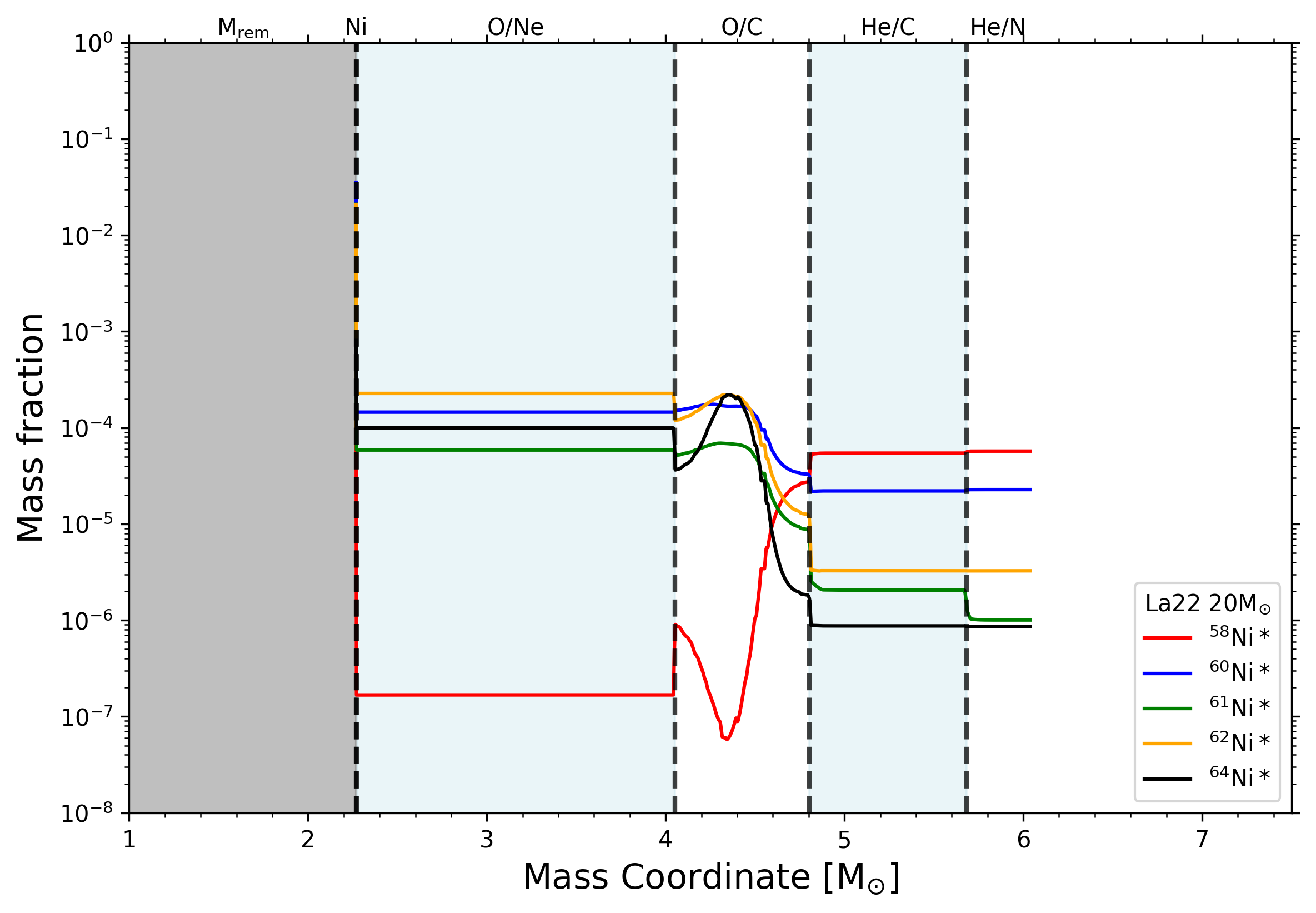}
\caption{Abundance profiles (in mass fraction [mass]) of all the stable Ni isotopes (different colors, as indicated in the legend box) as a function of the mass coordinate of the ejecta for the 20 \msun\ CCSN models of the 6 sets described in Section~\ref{sec:models}. Each set is indicated in the label box of each panel. Each model is separated into the different zone characterized by the most abundant burning products according to the classification of \citet{meyer95}.
Radiogenic production contributing to each isotope (according to Table~\ref{tab:Nitable}) is included in each abundance. No abundances reported in the database for the gray area at the lowest mass coordinates, given that this area is not part of the ejecta but of the remnant, compact object. The stellar surface is also not shown, as the composition remains constant in the H-rich region. Note that 
in Si18 and La22 the explosion is only followed up to a peak temperature of $\sim 1.5$ GK, therefore these models only include the zones up to the He/N. In all the other cases the envelope of the star is included in the simulation dominion and also the H zone is provided. 
}
\label{fig:Nispaghetti20}
\end{figure*}

Figure \ref{fig:Nispaghetti20} shows the Ni abundance profiles for the six 20 M$_{\odot}$ models, and Figure \ref{fig:Nislope20} shows the corresponding slope of $\varepsilon^{64}$Ni$_{(81)}$ versus $\varepsilon^{62}$Ni$_{(81)}$  internally normalized to $^{58}$Ni/$^{61}$Ni (see Eqs. \ref{eq:epsilon}, \ref{eq:intnorm}, \ref{eq:Q}, and \ref{eq:dil}). In these figures, the ejecta are divided in different zones, each of which is labeled according to the classification by \citet{meyer95} based on the most abundant nuclear-burning products present in each zone. For example, the zone labeled as ``Si'' is where explosive O burning occurred and Si is the most abundant element, while the zone labeled as ``He/C'' is where He burning partially occurred during the progenitor evolution and He and C are the most abundant elements. 

\subsection{Description of the stellar structure}

\begin{figure*}
     \includegraphics[width=0.5\linewidth]{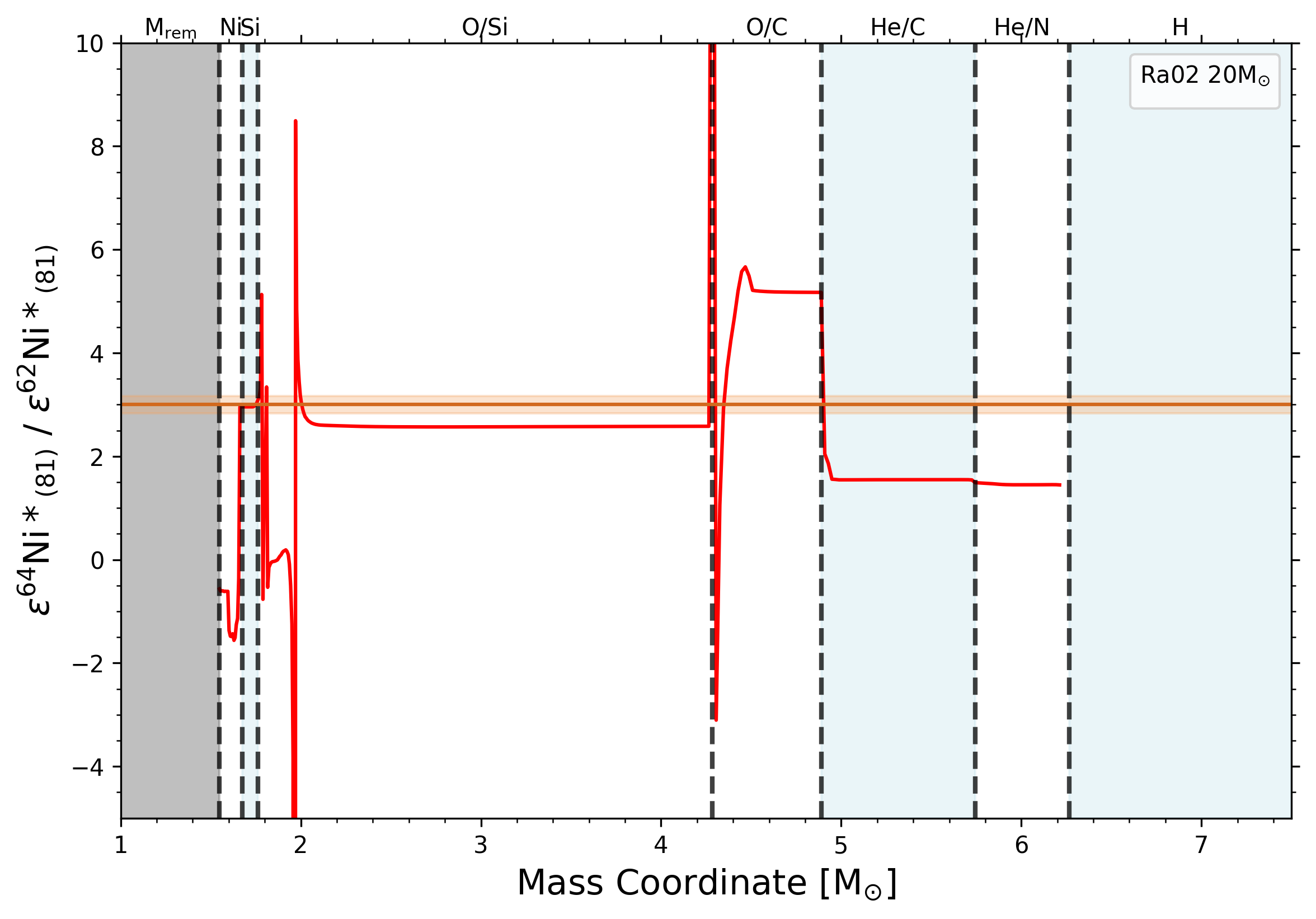}
     \includegraphics[width=0.5\linewidth]{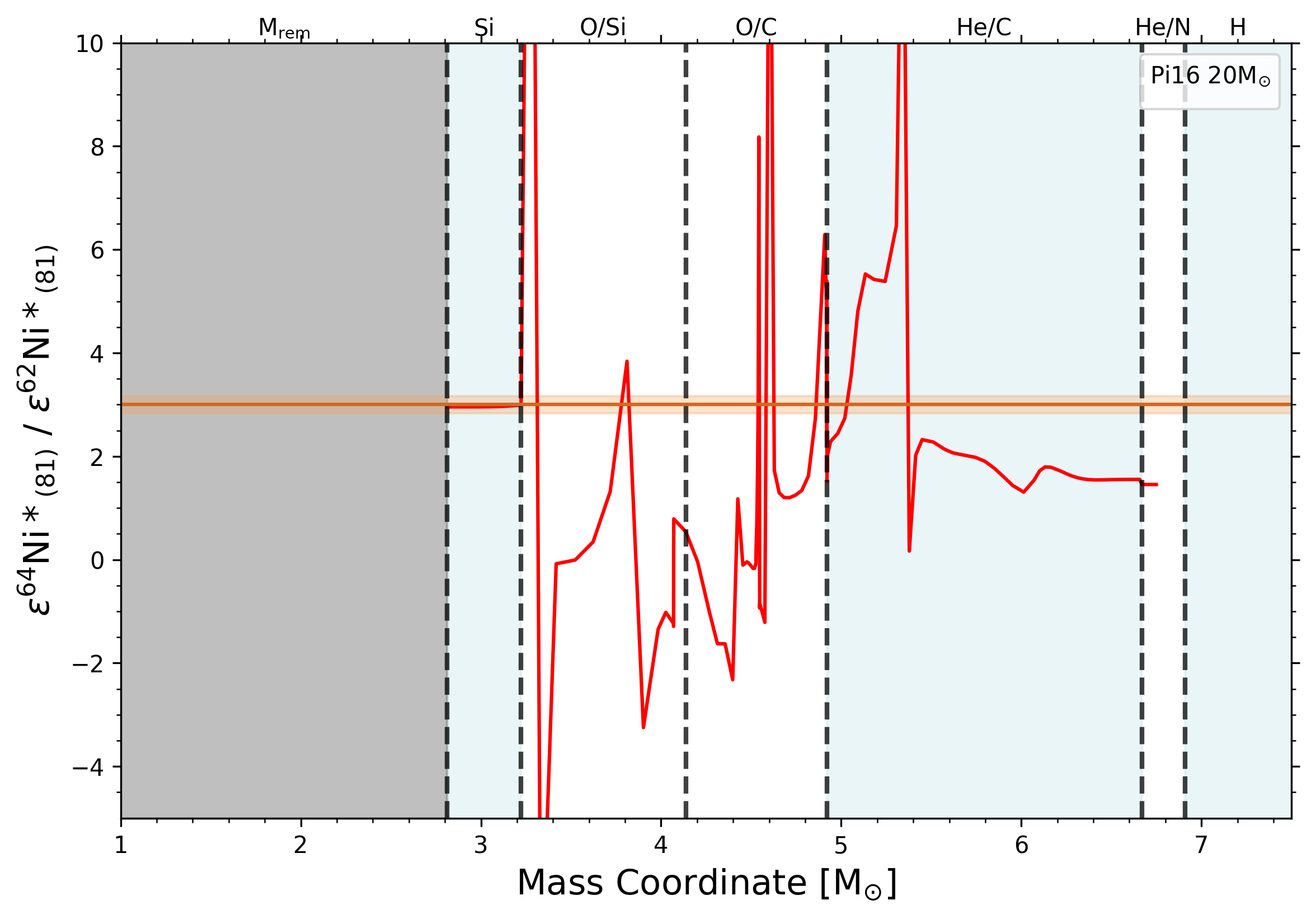}
     \includegraphics[width=0.5\linewidth]{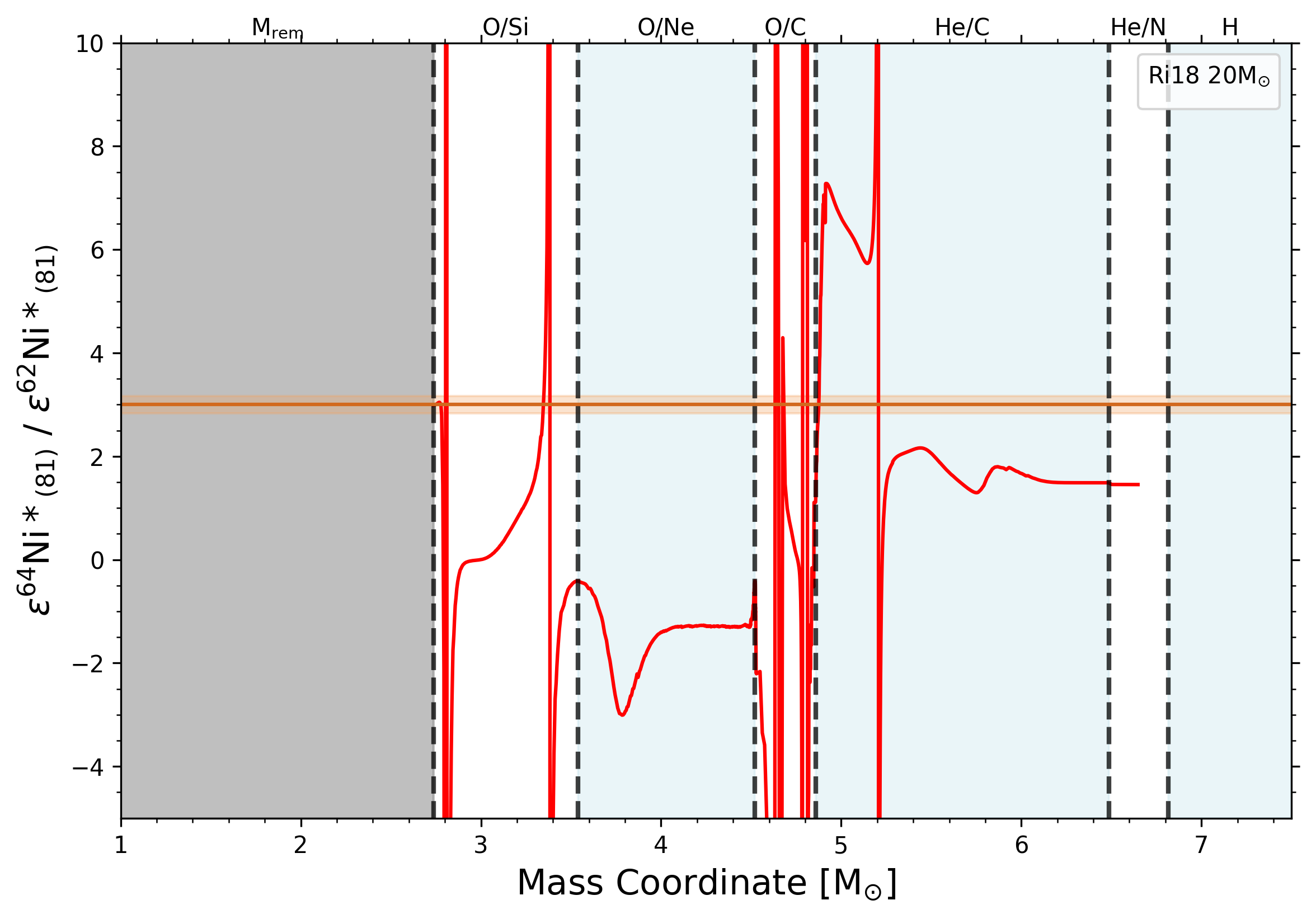}
     \includegraphics[width=0.5\linewidth]{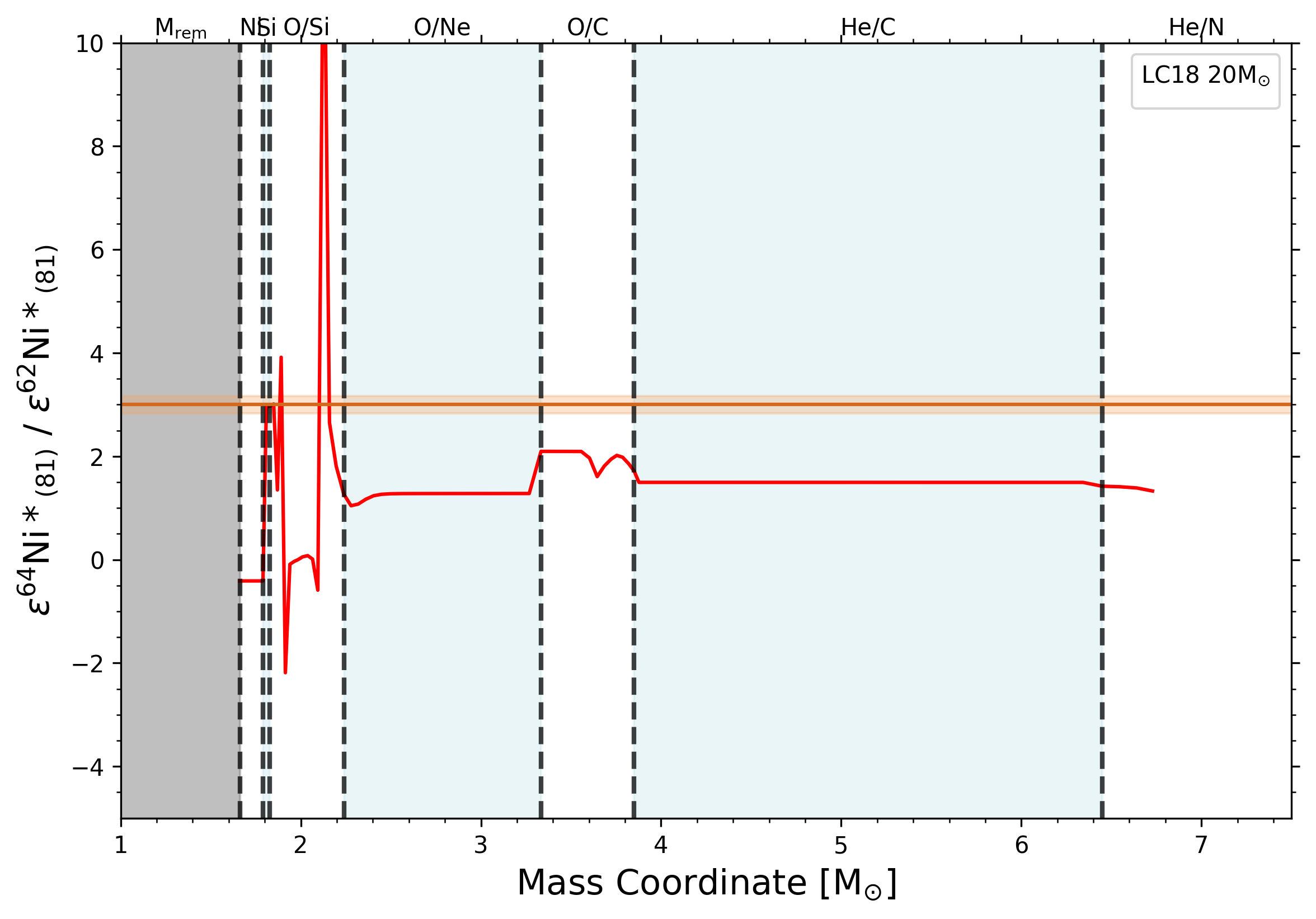}
     \includegraphics[width=0.5\linewidth]{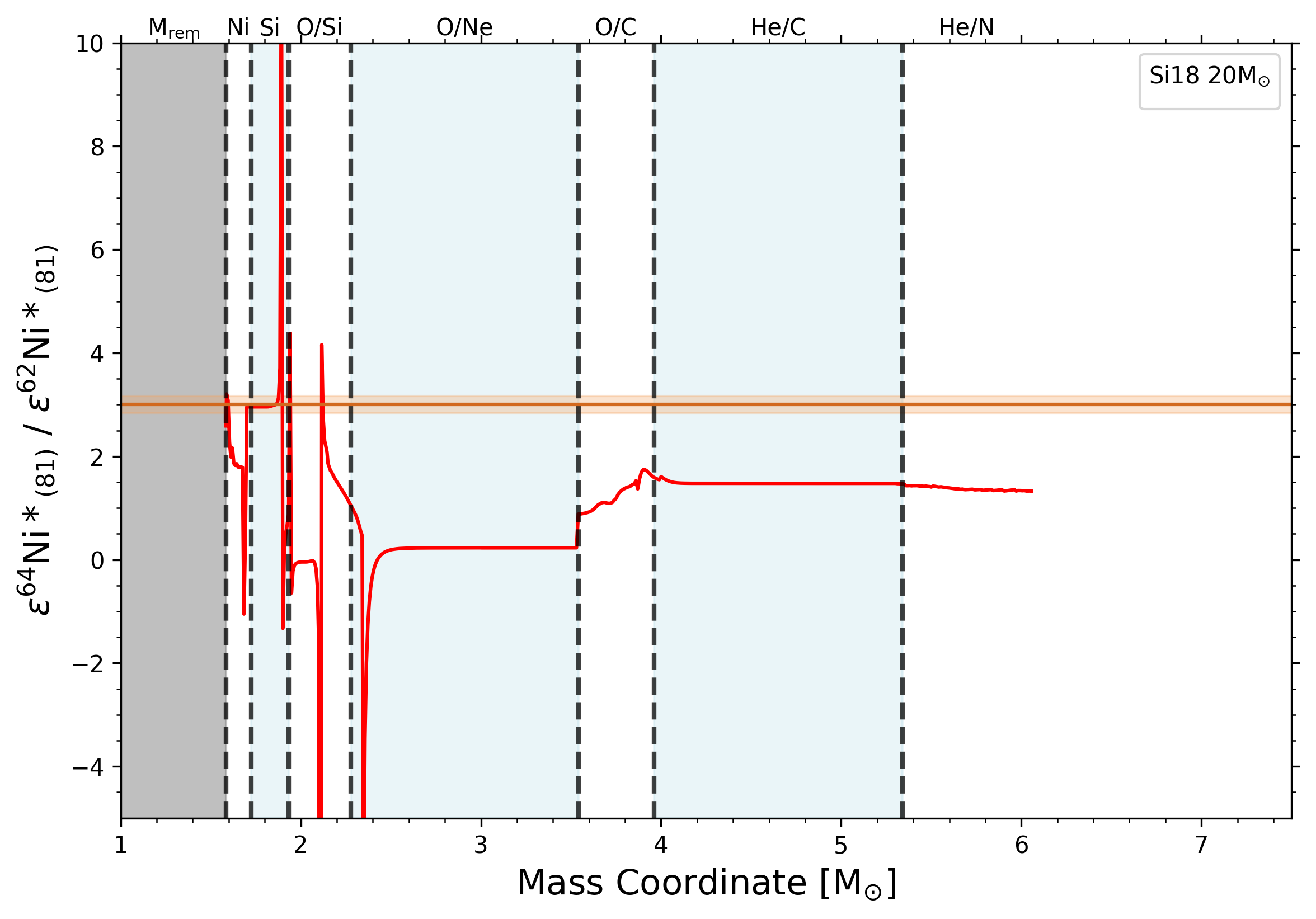}
     \includegraphics[width=0.5\linewidth]{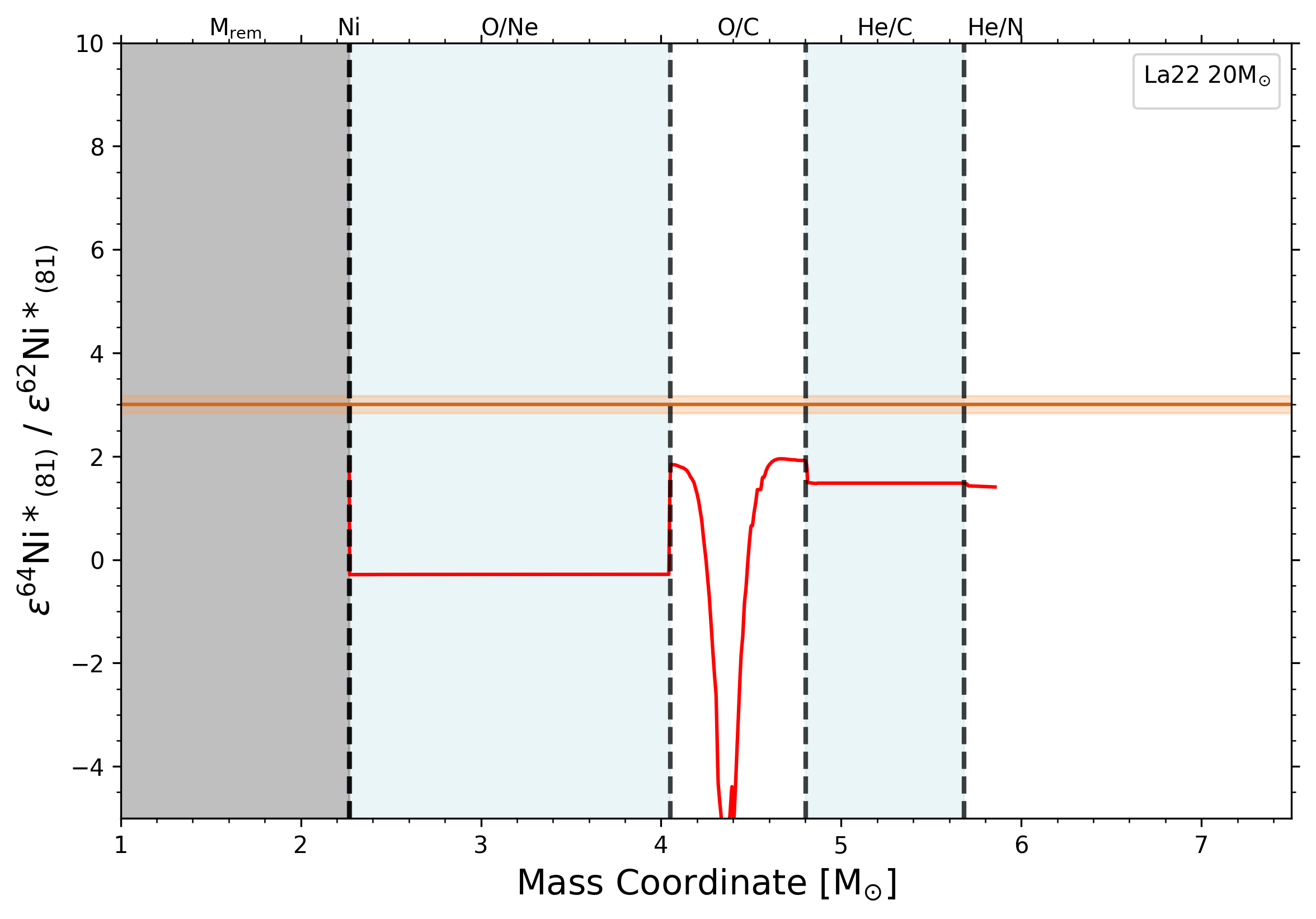}
\caption{The predicted slope of $\varepsilon^{64}$Ni$_{(81)}$ versus $\varepsilon^{62}$Ni$_{(81)}$ internally normalized to $^{58}$Ni/$^{61}$Ni, as indicated by the last digit of their masses in the subscript bracket (81) of the y-axis label, and where the asterisks indicate that all the radioactive isotopes are decayed into their corresponding stable daughter, 
calculated from the abundances shown in Figure~\ref{fig:Nispaghetti20}. The slope reported by \citet{steele12} of 3.003$\pm$0.166 from bulk meteoritic data is indicated for comparison as the horizontal orange line, with the shaded region representing the uncertainty.}
\label{fig:Nislope20}
\end{figure*}

The width and the location of the different zones defined above depend on multiple parameters, including the shell-burning processes during the pre-supernova evolution, the parametrization of the explosion, and the determination of the boundary between the ejecta and the remnant mass. The explosion primarily affects the definition and the width of Ni, Si, O/Si zones, and the location of the inner boundary of the O/Ne zone, while the outer boundary of the O/Ne zone, as well as the O/C, He/C, He/N, and H zones, are solely determined by the pre-supernova evolution. 
The inner boundary of the O/C zone defines the size of the CO core, which slightly changes after He burning, but still serves as a reliable indicator of the characteristics of the star \citep[e.g.,][]{arnett:85}.  
In our dataset, the inner border of the O/C zone varies in the range $\sim 3.4-4.5$ \msun, resulting in a variation of about 1 \msun\ among different models. Differences in pre-supernova evolution are also evident in the outer regions of the O/Ne zone. This region contains the ashes of the convective carbon shell, and variations in stellar evolution result in different chemical compositions, which can ultimately influence the slope plotted in Figure~\ref{fig:Nislope20} within this zone. While variations in the Ni isotopes are relatively small, the impact may be more significant for other elements. Another important difference is that the Ra02 model undergoes a C-O shell merger during pre-supernova evolution. This event leads to the removal of the O/Ne zone, as the merging of the carbon-burning shell with the oxygen-burning shell enhances the Si produced by the oxygen burning. As a result, this model exhibits only a large O/Si zone. Among the remaining models, this peculiar occurrence is observed only in the 15 \msun\ in the Ri18 set.

The different methods used to simulate the supernova explosion impact both the boundaries of the ejecta, as variations in fall-back prescriptions and mass-cut choices significantly influence the final remnant mass, while the explosion prescription can influence the upper boundary of the ejecta. 
For the 20 \msun\ models, the remnant mass varies between 1.55 \msun\ and 2.81 \msun. In the most extreme case, all the zones below the O/Ne are locked within the remnant. 

\begin{figure}
\begin{center}
\includegraphics[width=.48\textwidth]{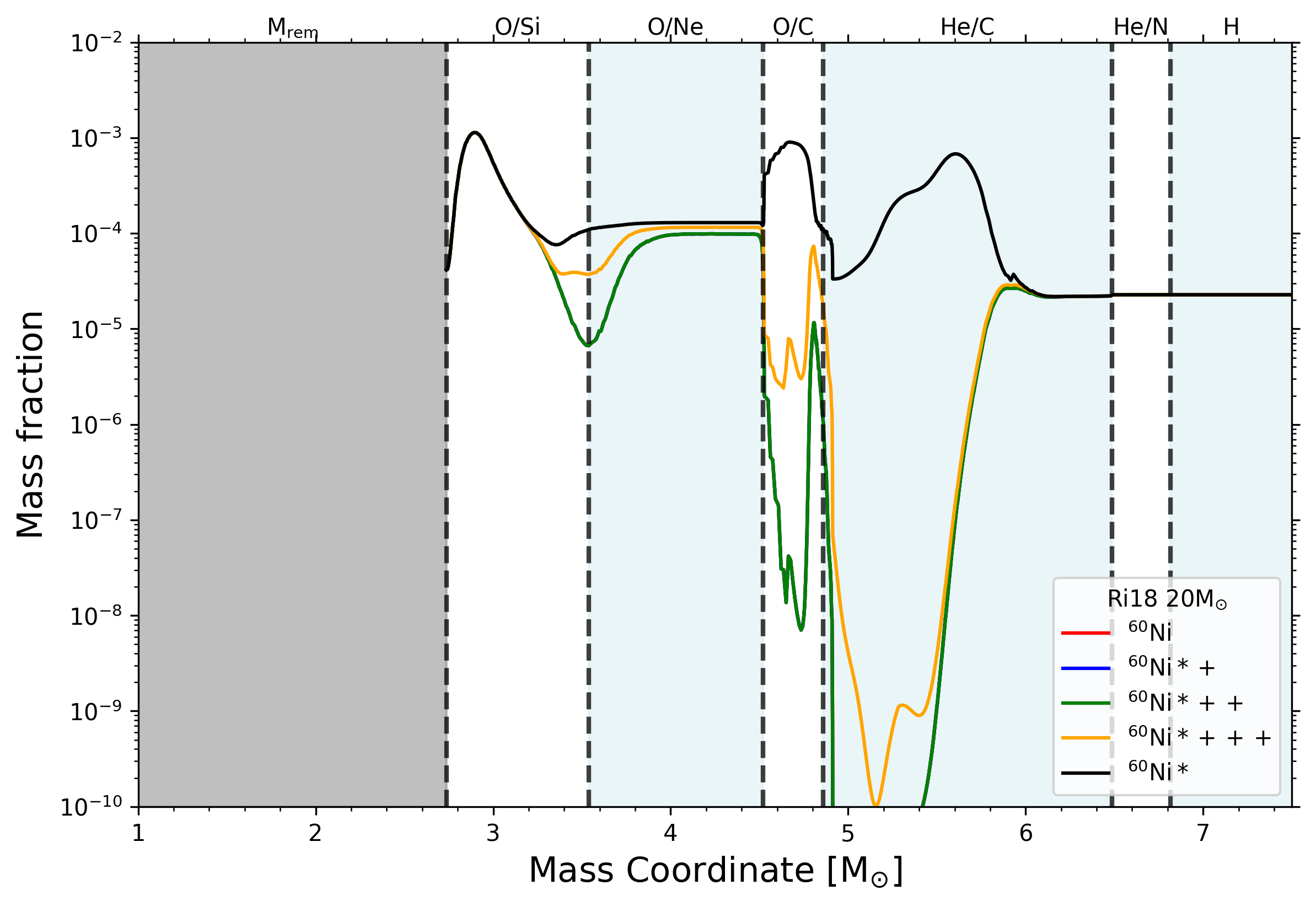}
\caption{\label{fig:ni60} 
Mass fraction [mass] of \iso{60}Ni as a function of the mass coordinate of the ejecta for the 20 \msun\ model of Ri18, with the different ejecta layers also indicated. The \iso{60}Ni abundance is calculated by progressively adding the abundances of the four radioactive isotopes listed in Table~\ref{tab:Nitable}. In the label box,  ``\iso{60}Ni'' corresponds to no radiogenic contributions, 
``\iso{60}Ni* +'' to the addition of \iso{60}Cu,   
``\iso{60}Ni* + +'' also of \iso{60}Zn, and 
``\iso{60}Ni* + + +'' also of \iso{60}Co. \iso{60}Ni* includes all the previous isotopes
plus \iso{60}Fe, therefore corresponding to the same line plotted in Figure~\ref{fig:Nispaghetti20}. Note that the red, blue and green lines overlap, which means that only \iso{60}Co and \iso{60}Fe would contribute to the abundance of \iso{60}Ni in this model.}
\label{fig:Ni60decayed_undecayed20}
\end{center}
\end{figure}
 
Some special events in the progenitor models might change the abundance profiles, also for the most abundant species, defying the standard zone classification used here following \citet{meyer95}. 
For instance, the Pi16 15 M$_{\odot}$ model developed a so-called C/Si zone in between the O/C zone and the He/C zone \citep[see also][]{pignatari13}. In several models calculated from the Pi16 25 M$_{\odot}$ model (not included in the SIMPLE current set), an O-rich O/nova zone develops close to the bottom of the He/C zone \citep[][]{pignatari2015, schofield:22}. We do not account for special cases in the current SIMPLE ejecta structure visualization package, and this will need to be updated to capture exotic zones produced by the simulations, especially as more models will be included in the future.

\subsection{Description of the Ni composition}

The Ni stable isotopes show a rich and diverse nucleosynthesis both in hydrostatic and CCSN conditions, resulting in a large range of variations in their abundances in the final ejecta. Explosive Si burning efficiently produces $^{58}$Ni, $^{61}$Ni and $^{62}$Ni, explosive O-burning mostly makes $^{58}$Ni, and the combination of the $slow$ neutron-capture ($s$) process during the progenitor evolution with the explosive neutron bursts in the convective C and He shells produces $^{60}$Ni, $^{61}$Ni, $^{62}$Ni, and $^{64}$Ni \citep[e.g.,][]{woosley02, pignatari:16}. 
To ensure the physical significance of the slopes plotted in Figure~\ref{fig:Nispaghetti20} it is useful to use  Figure~\ref{fig:Nispaghetti20} to verify for each region of the ejecta that the abundance of one or more isotopes is not depleted by orders of magnitude, or even approaching zero, relative to their corresponding initial, i.e., surface, abundances. In this case, while it would still be possible to calculate a slope, there would be no material to carry its signal.
While this consideration is not particularly relevant for Ni, which is generally abundantly produced in CCSNe, it must be carefully taken into account when examining elements with isotopes more scarcely produced, such as those heavier than Fe.

Figures~\ref{fig:Nispaghetti20} and \ref{fig:Nislope20} were generated by adding the abundances of all the radioactive isotopes listed in Table~\ref{tab:Nitable} to the stable Ni isotopes. While most of these isotopes decay in a less than a day, the decay chain $^{60}$Fe($\beta^-$)$^{60}$Co($\beta^-$)$^{60}$Ni has much longer half-lives (5.27 years for $^{60}$Co, and 2.62 million years for $^{60}$Fe). Therefore, these isotopes may not contribute to the abundance of $^{60}$Ni, depending on the dust formation time and the chemical behavior of these elements in different dust. For practical purposes, this is not relevant to the slopes shown in Figure~\ref{fig:Nislope20} because $^{60}$Ni is not included in the calculation of the $\varepsilon$ values considered there. 
In general, isotopes such as \iso{60}Ni with a potentially high radiogenic component require extra care when determining the origin of their variability to avoid confusion between the radiogenic and nucleosynthetic isotope variations. Here, we use this prominent example simply to illustrate how the evaluation of the impact of radioactive decay can easily be performed using the SIMPLE code by producing plots such as Figure~\ref{fig:ni60}, which shows which fraction of the abundance of $^{60}$Ni is due to the decay of its radioactive parents, and especially $^{60}$Fe, and in which region of the ejecta.

\begin{table}[h]
\begin{center}
\caption{\label{tab:decay} List of the radioactive isotopes whose abundances were added to those of the corresponding stable Ni isotopes to produce Figures~\ref{fig:Nispaghetti20} and \ref{fig:Nislope20}. Their decay mode and  
half-lives are also indicated, the latter in units of milliseconds (ms), seconds (s), hours (h), days (d), and million years (Myr) and with 1$\sigma$ error (from \url{https://www.nndc.bnl.gov/nudat3/}). 
}
\label{tab:Nitable}
\begin{tabular}{cccc} 
\hline
Stable & Unstable  & Half-life & Decay \\
isotope & isotope & & mode \\
\hline
\hline
$^{58}$Ni & $^{58}$Cu & (3.204 $\pm$ 0.007) s & $\beta^+$ \\
\hline
$^{60}$Ni & $^{60}$Cu & (23.7 $\pm$ 0.4) m & $\beta^+$ \\
       & $^{60}$Zn & (2.38 $\pm$ 0.05) m & $\beta^+$ \\
       & $^{60}$Co & (1925.28 $\pm$ 0.14) d & $\beta^-$ \\
       & $^{60}$Fe & (2.62 $\pm$ 0.04) Myr & $\beta^-$ \\
\hline
$^{61}$Ni & $^{61}$Cu & (3.339 $\pm$ 0.008) h & $\beta^+$ \\
       & $^{61}$Zn & (89.1 $\pm$ 0.2) s & $\beta^+$ \\
       & $^{61}$Co & (1.649 $\pm$ 0.005) h & $\beta^-$ \\
       & $^{61}$Fe & (5.98 $\pm$ 0.06) m & $\beta^-$ \\
\hline
$^{62}$Ni & $^{62}$Cu & (9.673 m $\pm$ 0.008) m & $\beta^+$ \\
       & $^{62}$Zn & (9.186 $\pm$ 0.013) h & $\beta^+$ \\
       & $^{62}$Co & (1.5 $\pm$ 0.4) m & $\beta^+$ \\
       & $^{62}$Fe & (68 $\pm$ 2) s & $\beta^-$ \\
\hline
$^{64}$Ni & $^{64}$Cu & (12.701 $\pm$ 0.002) h & $\beta^-$ \\
\hline
\end{tabular}
\end{center}
\end{table}

\begin{figure*}[ht!]
\begin{center}
\includegraphics[width=.9\textwidth]{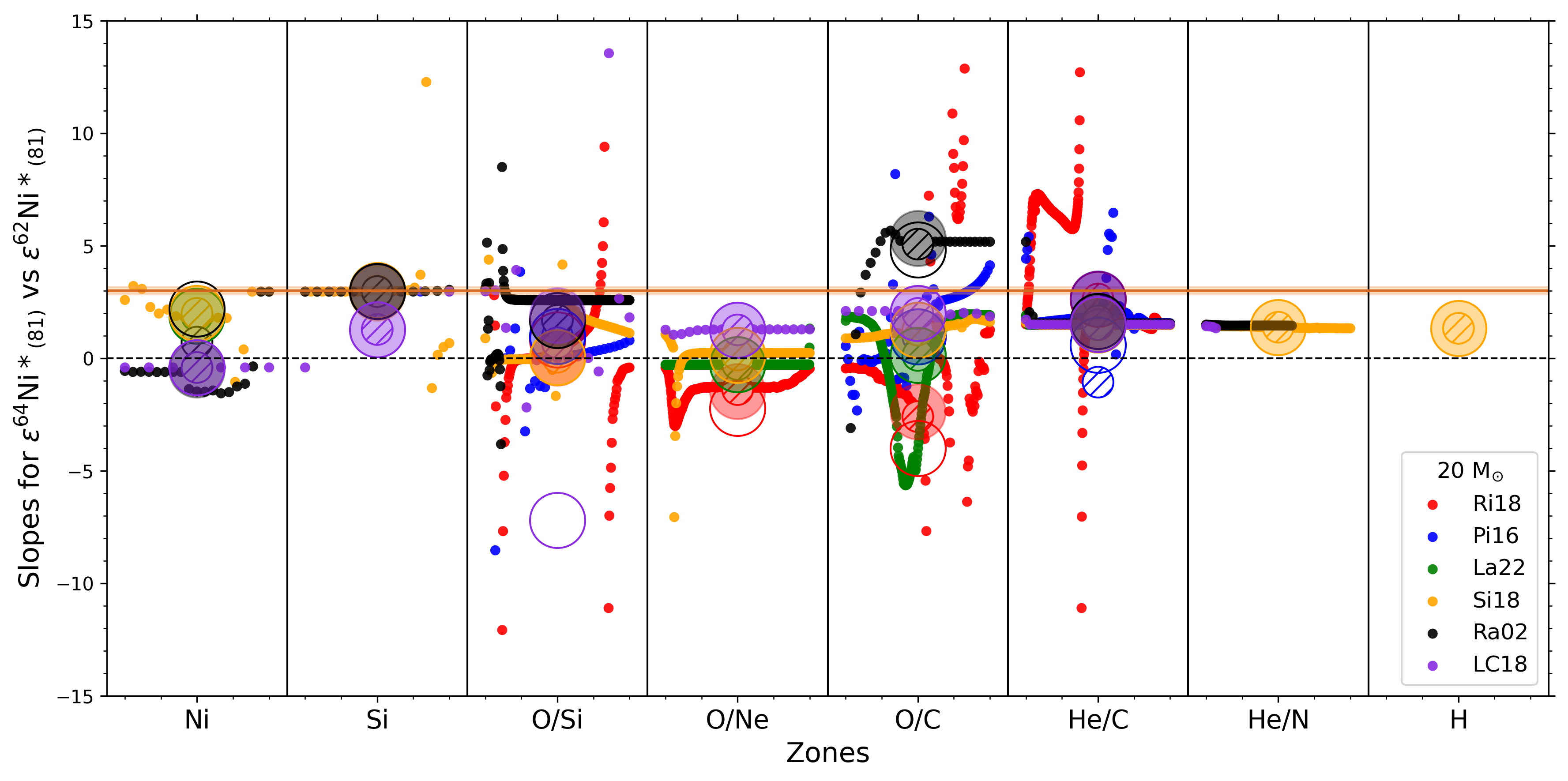}
\caption{Summary of all the predicted slope of $\varepsilon^{64}$Ni$_{(81)}$ versus $\varepsilon^{62}$Ni$_{(81)}$, internally normalized to $^{58}$Ni/$^{61}$Ni, as indicated by the last digit of their masses in the subscript bracket (81) of the y-axis label, and where the asterisks indicate that all the radioactive isotopes 
in each of the different zones of the 
6 sets considered, color-coded as indicated in the legend box. The small dots represent each local slope at a given mass coordinate, the large circles represent two different averages: weighted over the total mass included in each mass layer (filled circles), weighted over both the total mass and the total amount of Ni (in mass fraction) in each mass layer (empty circles), and as the previous but only including layers with Ni amounts varying within the 20th and the 80th percentile of the Ni distribution within the considered zone (i.e., with a statistical z-score lower than 1.25 - empty striped circles). 
\label{fig:zoneplot}}
\end{center}
\end{figure*}

The different explosion methods impact differently the Ni distribution in each of the zones defined above. The Ni-rich zone is the most internal zone of the ejecta, that contains material processed at very high temperatures ($\geq5$ GK); it is where the matter evolves at the nuclear statistical equilibrium (NSE). The different approaches used in the different codes to simulate this high temperature regime lead to a large difference in the Ni isotope distribution and in the slope. Conversely, all the models ejecting a Si zone have similar abundances of Ni isotopes (and therefore similar slopes), with a larger production of the neutron deficient nuclei $^{58}$Ni and $^{60}$Ni relative to the more neutron-rich isotopes, due to the high efficiency of photo-disintegration in this zone. The Ni outcomes of the O/Si zone depends on whether the star experienced a C-O shell merger and whether the explosion efficiently activated the 
$^{22}$Ne($\alpha$,n)\iso{25}Mg reaction, triggering an explosive neutron burst. As already mentioned, the C-O shell merger alters the external part of the O/Si zone, while the neutron burst enhances the production of $^{64}$Ni relative to $^{61}$Ni. In models where the neutron burst does not occur, these two nickel isotopes are co-produced in similar proportions. From the O/Ne zone, there is no signature of explosive nucleosynthesis, except in those models where the shock wave is able to be efficiently revived due to the density drop at the O/C zone interface. In these cases, some explosive He burning can occur, together with the explosive neutron burst. The models without explosive He burning instead, have only slight traces of the hydrostatic $slow$ neutron-capture process in the O/C and He/C zones.



A summary of the slopes from all the model sets for 20 \msun\ is shown in Figure \ref{fig:zoneplot}, organized by zone, including both the local slopes at each ejected mass coordinate and three different weighted averaged values. Single mass coordinates within the same zone can produce extremely variable results including outliers, some of which are located outside the figure boundary. The plotted averages are more representative of the Ni isotopic production in each zone. The figure shows the different signatures discussed above: neutron-capture processes in the He/C region, the broad production signatures in the O/C and O/Ne zones, and variations in the 
O/Si zone to several nucleosynthesis components present in these parts of the ejecta, from charged particle reactions in explosive Ne-burning and partial O-burning \citep[e.g.,][]{chieffi98}, to neutron-captures during explosive C burning \citep[][]{pignatari:16} and the $\gamma$-process \citep[e.g.,][]{roberti:23,roberti:24}. The He/N and the Si zones show the smallest variations in the slope, among all the different models. 

In the He/N zone, this is due to the fact that the temperatures are not high enough to efficiently affect the Ni isotopes. Here, the plotted results are driven by marginal production or destruction of some of the Ni isotopes. The small dispersion of the data at some ejected mass coordinates of the 15 \msun\ and 20 \msun\ is a good diagnostic of some active nucleosynthesis. However, it is unlikely that any realistic dilution factor 
would allow to preserve these such small signatures. It is also possible that the small anomaly seen in this region is driven by numerical errors introduced in the stellar calculations, where rounding numbers for the initial abundances and/or internal condition for mass conservation may generate small departures from the reference solar abundances. In general, we conclude that slopes obtained in the He/N zone should be considered with caution, except for a few light elements like CNO, which are affected by nucleosynthesis in this region. 
In the Si zone, instead, the Ni isotopes are strongly affected.  
Therefore, the fact that all the models able to eject a Si zone, independently from the progenitor mass and stellar set, provide us with a slope of $\sim$3 is a robust result. As described above and noted by \citet{steele12,cook21}, this result depends on the fact that in this region \iso{58}Ni is the strongly favored isotope, which inevitably results in a slope of $\sim$3, when internally normalized and calculated relative to the solar abundances. Therefore, we also conclude that when considering CCSNe as a source of the Ni anomaly, while the significant variability could provide a slope compatible with the observations in most of the ejecta, the most likely scenario is that they originated from an ejected Si zone. 



\section{Summary, discussion, conclusions, and future work}\label{sec:final}

We have presented the new python tool SIMPLE (Stellar Interpretation of Meteoritic Data and PLotting for Everyone), which comprises both software and a large collection of input data, as needed to compare stellar abundances in the ejecta of Fe-core-collapse supernovae (CCSNe) to meteoritic data. We provide input data from 18 models of solar metallicity, 3 different initial masses (15, 20, and 25 \msun), and 6 different sets (published in 2002, 2016, 2018, and 2022), to be able to reach more robust results than ever before. The code handles radioactive decay by allowing addition of the full abundances of unstable nuclei to their stable daughter nuclei. A more accurate implementation of the radioactive exponential law following a time evolution after the explosion will be implemented in the future. The code also allows calculation from stellar predictions of the correlation between isotopic ratios of different elements, with some basic implementation on the possible choice of the chemical fraction between two elements. While this was not the focus of our current presentation, we note that, with minimal modification, the code can be also be used for other applications, for instance, to calculate integrated abundance yields and/or to plot quantities useful for the comparison to stardust grains, from isotopic ratios to per mil variations relative to solar \citep[$\delta$-values, e.g.,][]{denhartogh:22}. 

As an illustrative case of the capabilities of SIMPLE, we have used the isotopes of Ni. We have described in detail the similarities and the peculiarities of the different models to provide a guideline for future studies on other elements. We confirm previous results \citep{steele12,cook21,hopp22} based on 4 models only, that the observed Ni variations in bulk meteorites can be robustly matched by material produced in the Si zone of the CCSN ejecta. 

SIMPLE can be easily extended to incorporate stellar abundance predictions from more models. 
Specifically for CCSNe, the current grid could be extended to results based on 3D prescriptions \citep[e.g.,][]{sieverding:23}, to more masses, for a more consistent comparison using the CO core rather than the initial mass (as discussed at the start of Section~\ref{sec:nickel}), as well as more explosion parameters, for example, by adding the other $>$60 models published in the La22 set. In general, it is debated which initial masses explode as CCSNe \citep[e.g.,][]{ugliano:12, ertl:16, sukhbold:16, boccioli:24} and while lower masses are more common, higher masses eject more mass, therefore, it will be useful to also extend the range of masses of the input models. 
Other sources to consider may be novae \citep[e.g.,][]{denissenkov:14} and other types of supernovae, such as thermonuclear supernovae \citep[e.g.,][]{keegans:23} and electron-capture supernovae \citep[][]{jones:19a}, as well as the nucleosynthetic outcomes of specific explosion-related processes, such as the neutrino winds, and for the stellar wind of massive stars \citep[e.g.,][]
{jones16,pignatari:16,ritter18set,brinkman21}. 
Predictions from intermediate- and low-mass stars in the asymptotic giant branch (AGB) phase \citep[e.g.,][]{karakas16, pignatari:16, battino16} could also be added as input to the code. 

Finally, to obtain a broader view of the compositions achievable in CCSNe, mixing of different mass layers (both adjacent and not) is another feature that will be useful to implement in the code, similarly to what has been recently done to study supernova light-curves \citep{fryer23,niblett25}. We aim for the SIMPLE code to represent a collaborative tool, welcome collaboration on its development, and plan to pursue its dissemination.


\begin{acknowledgments}
This work was supported by the European Union’s Horizon 2020 research and innovation programme (ChETEC-INFRA -- Project no. 101008324), the Lend\"ulet Program LP2023-10 of the Hungarian Academy of Sciences, the Hungarian NKFIH via K-project 138031 and NKKP Advanced grant 153697, and the IReNA network by NSF AccelNet (Grant No. OISE-1927130). GGB thanks the support provided by the undergraduate research assistant program of Konkoly Observatory. ML was also supported by the NKFIH excellence grant TKP2021-NKTA-64. 
MP thanks the 
support from the ERC Synergy Grant Programme (Geoastronomy, grant agreement number 101166936, Germany), 
Work by AS was performed under the auspices of the U.S. Department of Energy by Lawrence Livermore National Laboratory under contract DE-AC52-07NA27344. Lawrence Livermore National Security, LLC. This research has used the Astrohub online virtual research environment (https://astrohub.uvic.ca), developed and operated by the Computational Stellar Astrophysics group (http://csa.phys.uvic.ca) at the University of Victoria and hosted on the Computed Canada Arbutus Cloud at the University of Victoria. 
LR acknowledges the support from the PRIN URKA Grant Number \verb|prin_2022rjlwhn|.
\end{acknowledgments}

\bibliography{sample631,marialib1,marialib2,marialib3,apj-jour}

\end{document}